%% file: eccentricPE.tex
\documentclass[%
aps,
prd,
amssymb,
amsmath,
amsfonts,
eqsecnum,
showpacs,
nofootinbib,
floatfix,
twocolumn,
twoside,
a4paper,
superscriptaddress,
longbibliography
]{revtex4-2}

\input{UsePackages}
\usepackage{graphicx}
\usepackage{dcolumn}
\usepackage{bm}
\usepackage{tabularx}
\usepackage{multirow}
\usepackage{capt-of}
\usepackage{color}
\usepackage{url}
\usepackage[utf8]{inputenc}
\usepackage{placeins}

\usepackage{ulem}

\usepackage{graphicx}
\usepackage{booktabs}
\usepackage{adjustbox}
\usepackage{amsmath}

\usepackage{arydshln}
\usepackage{natbib}

\usepackage[nolist,nohyperlinks]{acronym}
\usepackage{xspace}
\usepackage[colorlinks]{hyperref}

\usepackage[caption=false]{subfig}

\usepackage{float}

\DeclareMathAlphabet{\mathbfsf}{\encodingdefault}{\sfdefault}{bx}{sl}

\newcommand{\be}{\begin{equation}}
\newcommand{\ee}{\end{equation}}
\newcommand{\bea}{\begin{eqnarray}}
\newcommand{\eea}{\end{eqnarray}}

\newcommand{\phT}{\textsc{IMRPhenomT}\xspace}
\newcommand{\phTE}{\textsc{IMRPhenomTEHM}\xspace}
\newcommand{\phTHM}{\textsc{IMRPhenomTHM}\xspace}
\newcommand{\phTPHM}{\textsc{IMRPhenomTPHM}\xspace}
\newcommand{\phXPHM}{\textsc{IMRPhenomXPHM}\xspace}

\newcommand{\NRSur}{\textsc{NRSur7dq4}\xspace}

\newcommand{\seobnrvfore}{\textsc{SEOBNRv4EHM}\xspace}
\newcommand{\seobnrvfivee}{\textsc{SEOBNRv5EHM}\xspace}
\newcommand{\gwNRHME}{\textsc{gwNRHME}\xspace}

\newcommand{\chieff}{\chi_\mathrm{eff}}
\newcommand{\chip}{\chi_\mathrm{p}}

\definecolor{dodgerblue}{HTML}{1E90FF}
\definecolor{viennared}{HTML}{DA0A14}
\definecolor{ctorange}{HTML}{FF6C0C}
\definecolor{granadagreen}{HTML}{078931}
\definecolor{wales}{HTML}{ff0038}
\definecolor{valenciacfred}{HTML}{ee3524}
\definecolor{barcelonafcgold}{HTML}{edbb00}
\definecolor{jam}{HTML}{A50B5E}
\definecolor{austriawien}{HTML}{441678}

\AtBeginDocument{%
  \hypersetup{
    citecolor=dodgerblue,
    linkcolor=dodgerblue,   
    urlcolor=dodgerblue}}

\newcommand{\UIB}{Departament de F\'isica, Universitat de les Illes Balears, IAC3 -- IEEC, Crta. Valldemossa km 7.5, E-07122 Palma, Spain}

\newcommand{\AEI}{Max Planck Institut für Gravitationsphysik (Albert Einstein Institut), Am M\"uhlenberg 1, Potsdam, Germany}

\newcommand{\ICE}
{Institut de Ci\`encies de l'Espai (ICE, CSIC), Campus UAB, Carrer de Can Magrans s/n, 08193 Cerdanyola del Vall\`es, Spain}

\newcommand{\Nikhef}
{Dutch National Institute for Subatomic Physics (Nikhef), Science Park 105, 1098 XG, Amsterdam, The Netherlands}

\newcommand{\ICS}{Astrophysik-Institut, Universit\"{a}t Z\"{u}rich, Winterthurerstrasse 190, 8057 Z\"{u}rich, Switzerland}

\allowdisplaybreaks[1]

\begin{document}

\title[ML]
{Eccentric or circular? A reanalysis of binary black hole gravitational wave events for orbital eccentricity signatures}

\author{Maria de Lluc Planas}
\affiliation{\UIB}

\author{Antoni Ramos-Buades}
\affiliation{\UIB}

\author{Cecilio García-Quirós}
\affiliation{\ICS}

\author{Héctor Estellés}
\affiliation{\AEI}

\author{Sascha Husa}
\affiliation{\ICE}
\affiliation{\UIB}

\author{Maria Haney}
\affiliation{\Nikhef}

\date{\today}

\begin{abstract}
We present a reanalysis of 17 gravitational-wave events detected
with Advanced LIGO and Advanced Virgo in their first three
observing runs, using the new \phTE model -- a
phenomenological time-domain multipolar waveform model for aligned-spin black-hole binaries in elliptical orbits with two eccentric parameters: eccentricity and mean anomaly. 
We also analyze all events with the underlying quasi-circular model \phTHM to study the impact of including eccentricity and compare the eccentric and quasi-circular binary hypotheses. 
The high computational efficiency of \phTE enables us to explore the impact of two different eccentricity priors --uniform and log-uniform-- as well as different sampler and data settings. 
We find evidence for eccentricity in two publicly available LVK events, GW200129 and GW200208\_22, with Bayes factors favoring the eccentric hypothesis over the quasi-circular aligned-spin scenario: $\log_{10}\mathcal{B}_{\mathrm{E/QC}}\in\left[1.30^{+0.15}_{-0.15}, 5.14^{+0.15}_{-0.15}\right]$ and $\log_{10}\mathcal{B}_{\mathrm{E/QC}}\in\left[0.49^{+0.08}_{-0.08}, 1.14^{+0.08}_{-0.08}\right]$, respectively. 
Additionally, the two high-mass events GW190701 and GW190929 exhibit potential eccentric features. For all four events, we conduct further analyses to study the impact of different sampler settings. We also investigate waveform systematics by exploring the support for spin precession using \phTPHM and \NRSur, offering new insights into the formation channels of detected binaries. 
Our results highlight the importance of considering eccentric waveform models in future observing runs, alongside precessing models, as they can help mitigate potential biases in parameter estimation studies. This will be particularly relevant with the expected increase in the diversity of the binary black hole population with new detectors.

\end{abstract}


\maketitle

\section{Introduction}
\label{sec:Introduction}

The first three observing runs of the Advanced Ligo and Advanced Virgo detectors ~\cite{Aasi_2015, Acernese_2015} have led to the 
detection of more than 90 gravitational wave (GW) events~\cite{gwtc1, gwtc2, gwtc21, gwtc3, Nitz_nov2021, Nitz_2023, Venumadhav:2019lyq, Olsen:2022pin,Wadekar:2023gea,Mehta:2023zlk}, all originating from compact object binaries. The majority of these events have been identified as binary black-hole (BBH) mergers, although several binary neutron star (BNS) and neutron star-black hole (NSBH) mergers have also been observed \cite{GW170817, GW190425, GW200105}. 
One of the key open questions in GW astrophysics is understanding the formation channels of these binaries.

The main uncertainty in the formation process stems from the fact that solely GW emission cannot drive widely separated binary stars into merging black holes (BHs)~\cite{Mandel:2018hfr}. Therefore, to explain the formation of the observed population of BBHs two main formation scenarios have been proposed in the literature, \textit{isolated evolution} and \textit{dynamical formation}. Within these two scenarios many formation sub-channels exists, for instance, common-envelope evolution \cite{Bethe:1998bn}, Population III stars \cite{Belczynski:2004gu}, or Zeipel-Kozai-Lidov oscillations \cite{zeipel1909,kozai1962secular,lidov1962evolution}, see Ref.~\cite{Mapelli:2020vfa} for a detailed review. 

BBHs formed from isolated binary evolutions are expected to circularize due to efficient angular momentum loss via gravitational radiation long before they enter the sensitive frequency band of ground-based detectors~\cite{PhysRev.136.B1224}.
Consequently, many existing GW waveform models for BBH mergers are constructed under the assumption of quasi-circular (QC) orbits~\cite{Bohe:2016gbl, Cotesta:2018fcv,Ossokine:2020kjp, Pompili_2023, Ramos-Buades:2023ehm, Gamboa:2024hli, Liu:2019jpg, Cao:2017ndf,Akcay:2020qrj, Nagar:2018plt, Nagar:2018zoe, Nagar:2020pcj, Nagar_2024, nagar2025, Gamba:2024cvy,Blackman:2017dfb, Varma:2019csw, Varma:2019vhw,Husa:2015iqa,Khan:2015jqa,London:2017bcn,Pratten:2020ceb,Garcia-Quiros:2020qpx,Estelles:2021gvs,Husa:2015iqa,Khan:2015jqa,London:2017bcn,Pratten:2020ceb,Garcia-Quiros:2020qpx,Estelles:2020twz, Estelles:2020osj, Estelles:2021gvs}. Moreover, the LIGO-Virgo-KAGRA Collaboration (LVK) ~\cite{gwtc1, gwtc2, gwtc21, gwtc3}, as well as independent studies~\cite{Nitz_nov2021, Nitz_2023, Venumadhav:2019lyq, Olsen:2022pin, Wadekar:2023gea, Mehta:2023zlk}, have found that the observed GW signals are largely consistent with QC binaries, and a variety of mechanisms have been proposed for this case, such as common envelope evolution~\cite{Tutukov_1993, Ivanova_2013, Mapelli_2018, Fragos_2019, Broekgaarden_2022}, chemically homogeneous evolution~\cite{deMink_2016, Mandel_2016, Marchant_2016}, stable mass accretion onto a black hole from its stellar companion~\cite{vanHeuvel_2017, Shao_2022} or ambient gas-driven fallback~\cite{Tagawa_2018}.

The alternative formation channel, \textit{dynamical formation}, can also lead to compact binary mergers within a Hubble time. In this scenario, the merger progenitors do not evolve together but instead form binaries dynamically through encounters in dense environments such as young star clusters, globular clusters, and galactic nuclei~\cite{Kremer_2020, Santoliquido_2020, McKernan_2018}. Frequent interactions in these regions can lead to compact objects swapping in and out of multiple binaries before ultimately merging. During a three-body interaction, gravitational binding energy from the initial binary is transferred to the ejected object, reducing the separation of the remaining binary and increasing the likelihood of a merger.
Future observations by the LVK and next-generation detectors, such as the Einstein Telescope (ET)~\cite{Maggiore:2019uih,Abac:2025saz}, Cosmic Explorer (CE)~\cite{Reitze:2019iox}, and space-based observatories like LISA~\cite{LISA:2022yao}, will provide access to a broader frequency range and significantly increase the number of detected events. This will enhance our ability to probe the properties of compact binaries, making it critical to develop robust methods for characterizing their evolutionary histories.

The formation history of a binary is imprinted on its component masses, spins, and orbital eccentricity. Identifying these signatures in GW observations is crucial for distinguishing between formation channels, yet it remains challenging due to the high-dimensional parameter space of compact binaries and the sensitivity limits of current detectors.
Orbital eccentricity in BBH mergers is a smoking gun of dynamical formation. Eccentricity can also help identify dark matter particles in ultralight boson clouds around BHs \cite{Boskovic:2024fga}. Since isolated binaries are expected to circularize before entering the LVK band, the detection of eccentricity in a BBH merger would strongly suggest a dynamical origin such as globular clusters or galactic nuclei~\cite{Zevin:2018kzq, Stone:2016ryd}, or through the Kozai-Lidov mechanism in triple systems~\cite{Wen:2002km, VanLandingham:2016ccd}. By studying the effects of eccentricity on GW signals, we can help quantify the fraction of BBH mergers arising from dynamical interactions.
Several studies have attempted to identify signatures of eccentricity in compact binary mergers. Early analyses primarily incorporated a single eccentricity parameter using the likelihood reweighting (importance sampling) method~\cite{Romero-Shaw:2019itr, Payne:2019wmy}, used in works such as \cite{Abbott:2019hdd, Romero-Shaw:2019itr, Nitz:2019spj, Romero-Shaw:2020thy, Gayathri:2020coq, Favata:2021vhw, OShea:2021faf, Romero-Shaw:2021lcs}. More recently, efforts have expanded to include two eccentricity parameters \cite{Ramos-Buades:2023yhy, Gupte:2024jfe, Gamboa:2024hli, Planas:2025feq}, as neglecting one of these parameters has been shown to introduce biases in parameter estimation (PE) studies \cite{Clarke:2022fma,Ramos-Buades:2023yhy}.

Most of the aforementioned studies were constrained by various limitations, such as sampling only on non-eccentric parameters and reweighting with eccentric models, restricting the analysis to a limited number of events, or relying on machine learning techniques to mitigate the high computational cost of eccentric waveform models with two eccentric parameters. In Ref.~\cite{Planas:2025feq}, we recently introduced \phTE, a time-domain phenomenological multipolar eccentric waveform model, which is built upon the aligned-spin QC model \phTHM and represents the most computationally efficient eccentric inspiral-merger-ringdown (IMR) model to date. 
There, we already presented PE studies of two real events: GW150914~\cite{GW150914, LIGOScientific:2016vlm}, the first detected GW signal from a BBH, and GW190521~\cite{GW190521}, a high total mass signal with potential signatures of eccentricity identified in the literature~\cite{Romero-Shaw:2020thy, Gamba_2023}.
These preliminary studies demonstrated the efficiency and accuracy of the model, recovering posteriors consistent with the literature. Neither of the events showed significant evidence for eccentricity. Specifically, the posteriors for GW190521 were uninformative, which aligns with the short duration of the signal and the fact that \phTE, like other state-of-the-art eccentric models, assumes circularization at merger. As a result, we would not expect to observe imprints of eccentricity in the merger-ringdown part of the signal, which was observed by the detectors. These findings are also reflected in recent studies on this event~\cite{Ramos-Buades:2023yhy,Gupte:2024jfe, Gamboa:2024imd}. 

In this work, we further analyze 17 BBH merger GW events identified in the literature as showing interesting features, such as precession, unequal masses, or potential eccentricity. We present results obtained using a uniform prior for eccentricity versus a logarithmic prior, and compare Bayes factors between the eccentric and QC hypotheses. We identify four events, GW190701, GW190929, GW200129, and GW200208\_22, with a preference for the eccentric hypothesis, consistent with the findings in the literature~\cite{Gupte:2024jfe, Romero-Shaw:2019itr}. 
Among these, GW200129 and GW200208\_22 show more support for eccentricity, while the evidence for eccentricity in GW190701 and GW190929 remains inconclusive. For these specific events, we conduct additional studies to examine the impact of various effects such as data treatment and sampler settings on the recovery of eccentricity. Moreover, we perform additional runs for these events using \phTPHM, the precessing time-domain phenomenological waveform model from the same \phT family, to study the preference for eccentricity over precession. Motivated by claims of precession for GW200129 \cite{Hannam_2022}, which were found only in precessing models that include mode asymmetries, we also conduct \NRSur~\cite{Varma:2019vhw} runs for this event.

This paper is structured as follows: In Sec.\ref{sec:methods}, we describe the methodology employed in this work. 
In Sec.~\ref{sec:results}, we present our results starting with an overview of all analyzed events, followed by a discussion of those that exhibit evidence of eccentricity. Finally, in Sec.~\ref{sec:conclusions}, we summarize our main findings and discuss potential avenues for future research.

\section{Methodology}\label{sec:methods}

In this section, we outline the methodology employed throughout this work. First, in Sec.~\ref{subsec:notation}, we revisit the notation and conventions used for parameterization. In Sec.~\ref{subsec:wfmodels},  we describe the waveform models used for the analysis: the eccentric \phTE model and its QC counterpart, \phTHM, as well as a brief introduction to the precessing models \phTPHM and \NRSur.
Next, in Sec.~\ref{subsec:GWdata}, we present the public data used for the analysis of the GW events. 
Finally, in Sec.~\ref{subsec:PEbasis}, we introduce our methods for performing Bayesian inference and its application to determining whether a detected GW event is more likely to be described by a BBH moving on an eccentric or QC orbit.

\subsection{Notation and conventions}\label{subsec:notation}
Throughout this paper, component masses are denoted by $m_i$ with $i=1,2$. We define the mass ratio $q = m_2/m_1 \leq 1$, and the symmetric mass ratio $\eta = {m_1 m_2}/{(m_1+m_2)^2}$. The chirp mass is given by $\mathcal{M}=\frac{(m_1m_2)^{3/5}}{M^{1/5}}$.
Note that masses refer in general to the detector's frame.
The $z$-component of the dimensionless spin magnitudes are denoted~$\chi_i=S_i^z/m_i^2$, which correspond to the projections of the dimensionless component spin vectors onto the orbital angular momentum.
We also report the effective-spin parameter $\chieff$~\cite{Etienne_2008, Ajith_2011, Santamaria_2010}, which captures the dominant nonprecessing-spin effects defined as 
\begin{equation}
    \chieff = \frac{m_1 \chi_1+m_2\chi_2}{M}.
    \label{eq:chieff}
\end{equation}
For the precessing runs performed with \phTPHM and \NRSur, we also provide the effective spin precession parameter $\chip$~\cite{Schmidt:2014iyl}. 
This parameter quantifies the in-plane spin effects, and corresponds to an approximate average over many precession cycles of the spins in the precessing orbital plane, given by
\begin{equation}
    \chip=\frac{S_{\rm{p}}}{A_1m_1^2}.
    \label{eq:chip}
\end{equation}
Here, $S_{\rm{p}}$ is the average spin magnitude, given by
\begin{equation}
    S_{\mathrm{p}}=\max(A_1S_{1\perp}, A_2S_{2\perp}),
\end{equation}
where $A_1=2+3/2q$, and $A_2=2+3q/2$.

Comparing eccentric parameters across different waveform models requires additional post-processing due to the gauge dependence of eccentricity in General Relativity. The \phTE model, for instance, allows for two different gauges for the eccentric parameters, denoted as $e^{\rm PN}$ and $e^{\rm EOB}$ 
corresponding to modified-harmonic PN (MH) and alternatively Effective-One-Body (EOB, \textit{default}) coordinates (see Sec.~III C in Ref.~\cite{Planas:2025feq} for more details). However, gauge-invariant approaches to define eccentricity have been proposed, based on waveform modulations~\cite{Shaikh:2023ypz, Islam:2025oiv} and catastrophe theory~\cite{Boschini:2024scu}. In this work, we adopt the waveform-based eccentricity definition from Ref.~\cite{Ramos-Buades:2022lgf}, denoted as $e^{\rm GW}$, and implemented in the \texttt{gw\_eccentricity} python package \cite{Shaikh:2023ypz}.
This definition is computed by measuring the instantaneous angular GW frequency of the $(2,2)$ spherical harmonic mode, $\omega_{22}$, at the pericenters and apocenters of the orbits,
\begin{align}
    \label{eq:egw} e^{\rm GW}&=\cos(\psi/3)-\sqrt{3}\sin(\psi/3), \\
    \label{eq:psi} \psi & =\arctan\left(\frac{1-e^2_{\omega_{22}}}{2e_{\omega_{22}}}\right), \\
   \label{eq:eomega22} e_{\omega_{22}} &=\frac{\omega^{1/2}_{22,p}-\omega^{1/2}_{22,a}}{\omega^{1/2}_{22,p}+\omega^{1/2}_{22,a}    },
\end{align}
where $\omega^{1/2}_{22,p}$ and $\omega^{1/2}_{22,a}$ are the GW frequency of the $(2,2)$-mode at the periastra and apastra, respectively.
Similarly, the GW mean anomaly is given by \cite{schmidt2017,Shaikh:2023ypz}
\begin{equation}
l^{\rm GW}(t)=2\pi\frac{t-t^p_i}{t^{p}_{i+1}+t^{p}_{i}
},
\label{eq:lgw}
\end{equation}
where $t_i^p$ is the time of the $i-$th periastron passage measured from $\omega_{22}$.

\subsection{Waveform models}\label{subsec:wfmodels}

State-of-the-art gravitational IMR waveform models accurately describe BBHs in QC orbits, including also the effects of black holes spins. 
These models fall into three main families: the EOB formalism~\cite{Buonanno:1998gg, Buonanno:2000ef}, including SEOBNR~\cite{Bohe:2016gbl, Cotesta:2018fcv,Ossokine:2020kjp, Pompili_2023, Ramos-Buades:2023ehm, Gamboa:2024hli, Liu:2019jpg, Cao:2017ndf} and TEOBResumS models~\cite{Akcay:2020qrj, Nagar:2018plt, Nagar:2018zoe, Nagar:2020pcj, Nagar_2024, nagar2025, Gamba:2024cvy}; NRSurrogate models~\cite{Blackman:2017dfb, Varma:2019csw, Varma:2019vhw, Islam:2022laz, Rink:2024swg}, which interpolate between NR datasets; and the IMRPhenom approach~\cite{Husa:2015iqa,Khan:2015jqa,London:2017bcn,Pratten:2020ceb,Garcia-Quiros:2020qpx,Estelles:2021gvs,Husa:2015iqa,Khan:2015jqa,London:2017bcn,Pratten:2020ceb,Garcia-Quiros:2020qpx,Estelles:2020twz, Estelles:2020osj, Estelles:2021gvs}, known for its computational efficiency. 

For the QC aligned-spin subspace, the models are calibrated to NR simulations and demonstrate strong agreement in the region where NR data is available~\cite{Varma:2019csw, Cotesta:2018fcv, London:2017bcn, Garcia-Quiros:2020qpx, Estelles:2020twz}. 
Here, we employ the QC \phTHM model~\cite{Estelles:2020twz}, an extension of the time-domain IMR phenomenological model~\phT~\cite{Estelles:2020osj} to also include the subdominant spherical harmonics $(l,m)= \{(2,\pm 1), (3,\pm 3), (4,\pm 4), (5,\pm 5)\}$ in addition to the $(2,\pm 2)$-modes. The model is calibrated to SEOB-NR hybrids, up to mass ratio $18$, and to Teukolsky-based solutions in the extreme mass ratio limit. This results in an accurate and efficient model, widely used in GW astronomy \cite{Colleoni:2020tgc, Mateu-Lucena:2021siq, Estelles:2021jnz}.

Modeling precessing-spin binaries presents challenges due to the more complex waveform morphology and the larger parameter space involved, affecting all different modeling approaches~\cite{Ramos-Buades:2023ehm, Estelles:2021gvs, Varma:2019vhw, Eleanor_2021, Thompson:2023ase}. 
In this work, we rely on both the \phTPHM and \NRSur models.
The state-of-the-art precessing time-domain model, \phTPHM~\cite{Estelles:2021gvs}, uses approximations to model precession without NR calibration. While this increases its applicability across the parameter space, it comes at the cost of reduced accuracy. It includes the same multipoles as the underlying aligned-spin model \phTHM, and the default implementation used in this work uses a numerical integration of the Euler angles to parameterize the binary evolution. 
One limitation of non-calibrated precessing models is the neglect of $m$-mode asymmetries, which can lead to biases in PE studies, as demonstrated in Ref.~\cite{Hannam_2022, Borchers_2024}.
To account for the limitations of this model, we also use the
\NRSur~\cite{Varma:2019vhw} model, a more accurate precessing QC model, to study the precessing hypothesis for GW200129 claimed in other studies~\cite{Hannam_2022}. A key advantage of \NRSur is that it includes mode-$m$ asymmetries, which arise naturally from the numerical relativity simulations used in its calibration.
\NRSur is however limited in its mass ratio, spin magnitudes, and waveform length coverage.

Similar challenges arise in modeling eccentric binaries. The increased parameter space and the limited availability of NR eccentric simulations complicate both the development and validation of accurate waveform models. 
A common approach in the literature is the assumption of circularization before merger, which has been used to construct hybrid IMR waveforms by combining EOB/PN-inspiral waveforms with merger and ringdown signals from NR or the EOB formalism~\cite{Ramos-Buades:2019uvh, Hinder:2017sxy, Huerta:2017kez}. Significant progress has been made within the EOB framework~\cite{Placidi:2021rkh, Khalil:2021txt, Albanesi:2021rby, Nagar:2021xnh, Nagar:2021gss, Liu:2021pkr, Chiaramello:2020ehz, Liu:2019jpg, Cao:2017ndf, Hinderer:2017jcs, Ramos-Buades:2021adz, Carullo:2023kvj}, including recent efforts to develop generic IMR waveforms~\cite{Gamba:2024cvy, Liu_2023}.
Surrogate modeling has also been explored, but due to the scarcity of NR eccentric simulations, current models are reduced to comparable-mass and non-spinning binaries~\cite{Islam:2021mha, Islam:2024zqo}. The recently developed \gwNRHME framework can convert multi-modal QC waveforms into multi-modal eccentric waveforms, for a given quadrupolar eccentric waveform from a non-spinning system~\cite{Islam:2024bza, Islam:2024rhm,Islam:2024zqo, Islam:2025oiv}.

In this work, we use \phTE, the first eccentric model of the \phT family, which extends the QC \phTHM~\cite{Estelles:2020osj, Estelles:2020twz} model to eccentric binaries.
It includes the full 3PN orbit-averaged dynamics, accounting for both non-spinning and spinning corrections~\cite{Henry:2023tka}, implemented in both modified harmonic PN (referred to as PN) and EOB coordinates (default option). Additionally, the waveform modes of \phTHM are modified by incorporating non-spinning eccentric corrections up to 3PN for the non-spinning contributions~\cite{Ebersold:2019kdc}, and up to 2PN nonprecessing-spin eccentric corrections~\cite{ Henry:2023tka}, formulated in MH coordinates and expanded in eccentricity up to $\mathcal{O}(e^6)$. This limits the applicability of the model to BBHs with aligned spins and eccentricities below $e=0.4$ at an orbit-averaged $(2,2)$-mode frequency of 10 Hz.

\subsection{GW real events data}\label{subsec:GWdata}
We analyze 17 GW events from the public strain data available at the Gravitational Wave Open Science Center (GWOSC)~\cite{gwosc12,gwosc3}, using power spectral densities (PSDs) and calibration uncertainties provided by the LVK Collaboration. 
These events span the three observing runs (O1-O3) from mid-2015 to March 2020. During O1, the LIGO detectors were the only ones operational~\cite{LIGOScientific:2016emj}, and Virgo joined at the end of O2 in 2017~\cite{gwtc1}.

The selection of events in this work is based on particular and interesting features identified in previous studies. 
From GWTC1~\cite{gwtc1}, we analyze most of the detecting events, excluding those with very low total mass, such as GW170817~\cite{GW170817} (a binary neutron star merger) and GW170608.
From GWTC2~\cite{gwtc2, gwtc21}, we include GW190412~\cite{GW190412} due to its asymmetric masses and the impact of higher-order multipoles (HMs); GW190620, which has been argued to support a non-eccentricity explanation~\cite{Romero-Shaw:2021lcs}; GW190701 and GW190706, both of which showed potential signs of eccentricity~\cite{Gupte:2024jfe}; GW190814~\cite{GW190814}, the most asymmetric system yet measured with GWs, with a secondary component being either the lightest BH or the heaviest NS ever discovered in a double compact-object system; and GW190828 and GW190929.
Finally, from GWTC3~\cite{gwtc3}, we revisit GW200208\_22, which showed potential eccentricity signatures~\cite{Gupte:2024jfe, Romero-Shaw:2022xko}, and GW200129, initially reported as the first precessing binary detection~\cite{Hannam_2022} but later found to be affected by glitches~\cite{Payne_2022, Davis:2022ird}.

Several studies have been conducted to mitigate the known glitches in the event GW200129, including the use of machine learning techniques~\cite{Macas_2024, Malz_2024}. In this work, we use the frame files provided in the public release of Ref.~\cite{Payne_2022} to assess the impact of glitch subtraction on the eccentricity posteriors.
Specifically, Ref~\cite{Payne_2022} employs the \texttt{gw\_subtract}~\cite{Davis:2018yrz, Davis:2022ird} and \texttt{BayesWave}~\cite{Cornish:2014kda} glitch mitigation techniques.
The \texttt{gw\_subtract} method estimates and removes instrumental noise by using an auxiliary sensor at LIGO Livingston that monitors the noise source. The transfer function between this sensor and the main strain data channel is computed over a long period, allowing for an estimate of the noise contribution, which is then subtracted from the strain data. The effectiveness of this method depends on the accuracy of both the sensor and the transfer function estimate. This is the preferred technique employed in most analyses of this event~\cite{Hannam_2022,Gupte:2024jfe,Maggio_2023} and is also used in the GWTC-3 LVK analysis~\cite{gwtc3}.
On the other hand, \texttt{BayesWave} is a data analysis algorithm that models astrophysical signals, instrumental glitches, and Gaussian noise using a trans-dimensional Reversible-Jump Markov Chain Monte Carlo method~\cite{10.1093/biomet/82.4.711}. It represents the signal with a waveform model and incoherent non-Gaussian noise with sine-Gaussian wavelets, while simultaneously modeling the PSD with a combination of cubic splines and Lorentzians. The algorithm infers posterior distributions for the signal, glitch, and PSD, from which one can draw realizations of the glitch and subtract them from the detector strain. 

\subsection{Parameter Estimation: Bayesian Inference}\label{subsec:PEbasis}
Bayesian inference is the standard statistical framework in GW astronomy to measure the source properties of the detected GW events. 
The Bayes' theorem allows to obtain a distribution of the set of parameters $\lambda$ which characterize the GW source, provided that the observed GW signal in the detector data $d$ can be described by a theoretical signal model $h_M(t;\lambda)$.
The posterior probability distribution on $\lambda$ given the signal model $h_M$, $P(\lambda|d, h_M)$, is given by
\begin{equation}
    P(\lambda|d, h_M)=\frac{P(d|\lambda,h_M)  P(\lambda| h_M) }{P(d|h_M)}.
    \label{eq:bayes}
\end{equation}
Here, $P(d|\lambda,h_M)$ is the likelihood function, $P(\lambda| h_M)$ is the prior probability distribution, and $P(d|h_M)$ is the so called evidence of the model hypothesis $h_M$, normally represented by $\mathcal{Z}$, and defined as
\begin{equation}
    \mathcal{Z}=P(d|h_M)=\int \mathrm{d}\lambda P(d|\lambda,h_M)P(\lambda|h_M).
    \label{eq:evidence}
\end{equation}
In this work, we use \texttt{bilby}~\cite{bilby, bilby_pipe_paper} to compute the posterior distributions. By default, we use the nested sampler \texttt{dynesty}~\cite{dynesty}, fixing the number of autocorrelation times to use before a point to \texttt{naccept=60} and the number of live points to \texttt{nlive=1000}.

For a detector with stationary, Gaussian noise, the likelihood function is
\begin{equation}
    P(d|\lambda,h_M)\propto \exp\left[-\frac{1}{2}\langle d- h_M(\lambda)|d-h_M(\lambda)\rangle\right],
\end{equation}
where the noise-weighted inner product is defined as
\begin{equation}
    \langle A|B\rangle=2\Re\int_{f_{\mathrm{low}}}^{f_{\mathrm{high}}}\mathrm{d}f\frac{\tilde{A}^*(f)\tilde{B}(f)+\tilde{A}(f)\tilde{B}^*(f)}{S_n(f)}.
\end{equation}
Here, tildes denote Fourier transforms, asterisks denote complex conjugates, and $S_n(f)$ is the one-sided PSD of the detector. The integration limits $[f_{\rm low}, f_{\rm high}]$ define the detector's bandwidth, with $f_{\rm low}=20$ Hz in all cases, while $f_{\rm high}$ varies per run according to the official LVK reanalysis~\cite{gwtc1, gwtc2, gwtc21, gwtc3}. For multiple detectors, we assume uncorrelated noise, so the network likelihood is the product of individual likelihoods.

Bayesian-inferred posteriors provide a useful tool for comparing different models. Specifically, they allow us to estimate the preference for an eccentric model, E, over a QC model, by comparing their probabilities given the detector data:
\begin{align}
    \mathcal{O}_{\rm E/QC}& =\frac{P(h_{\rm{E}}|d)}{P(h_{\rm QC}|d)}=\frac{p(h_{\rm E})p(d|h_{\rm E})}{p(h_{\rm QC})p(d|h_{\rm{QC}} )}=\\
    & =\frac{p(h_{\rm E})}{p(h_{\rm QC})}\frac{\mathcal{Z}_{\rm E}}{\mathcal{Z}_{\rm {QC}}}=\frac{p(h_{\rm E})}{p(h_{\rm QC})}\mathcal{B}_{\rm E/QC}.
\end{align}
The first ratio represents our \textit{prior belief} about the occurrence of eccentric versus QC events in the universe. This prior knowledge can be inferred from previous GW observing runs, which provide event rate estimates for different populations. The second ratio, $\mathcal{B}_{\rm E/QC}$, is the Bayes factor, which quantifies the relative evidence for eccentricity compared to the QC assumption based on the observed data.
In this work, we compute the log-10 Bayes factor, $\log_{10}\mathcal{B}_{\mathrm{E}/\mathrm{QC}}$, which is positive when the eccentric model is preferred. Following Jeffreys' scale of evidence~\cite{jeffreys1961theory}, we consider $\log_{10}\mathcal{B}_{\mathrm{E}/\mathrm{QC}}>1$ as a threshold for strong support in favor of eccentricity.

We adopt priors on the inverse mass ratio ($1/q$) and chirp mass ($\mathcal{M}$) to ensure a uniform distribution in the component masses. 
For the spin components $\chi_i$, we use priors corresponding to the projections of a uniform and isotropic spin distribution along a direction perpendicular to the binary's orbital plane~\cite{PhysRevD.91.042003}.
For the luminosity distance $d_L$, we follow the simple prior proportional to $d_L^2$~\cite{gwtc1, gwtc2, gwtc21, gwtc3}, which distributes mergers uniformly through a stationary Euclidean universe. 
We set the starting frequency for the waveform generation at 10 Hz to ensure that the $l \leq 4$ modes remain within the analysis frequency band. Higher modes with $m > 2$ have higher frequency content over the same time interval, so the starting frequency must be adjusted based on the highest $m$-mode included in the analysis. While we include all available modes of the \phTE model, we limit the starting frequency to the $l = 4$ mode. This is because the additional time required for conditioning, along with the small effect of the $(5,5)$ mode at lower frequencies, make further reduction unnecessary for the waveform.
Regarding the two extra priors for eccentric models, we set a uniform distribution for the mean anomaly at the reference frequency of 10 Hz, $l_{\text{10\ Hz}}\in[0,2\pi]$. 
For the eccentricity at the reference time, we consider two choices: \textit{a)} a uniform prior, $e_{\text{10\ Hz}}\in[0,0.5]$, and \textit{b)} a log-uniform prior with bounds $e_{\text{10\ Hz}}\in[10^{-4},0.5]$. 
We analyze both priors to compare the resulting posterior distributions. The log-uniform prior is often used to express ignorance of the order of magnitude of the eccentricity, and is consistent with the astrophysical expectations that
typical eccentricities will be very small. 
However, using a uniform prior avoids imposing a lower bound on the log-uniform prior, which could significantly affect Bayes factors by assigning excessive weight to low eccentricities in the evidence integral (Eq.~\eqref{eq:evidence}).
The remaining priors, including the extrinsic parameters and the binary's orbital phase $\varphi$, are the same as in Ref.~\cite{gwtc1}. 

\begin{figure*}
    \centering
    \includegraphics[width=1\linewidth]{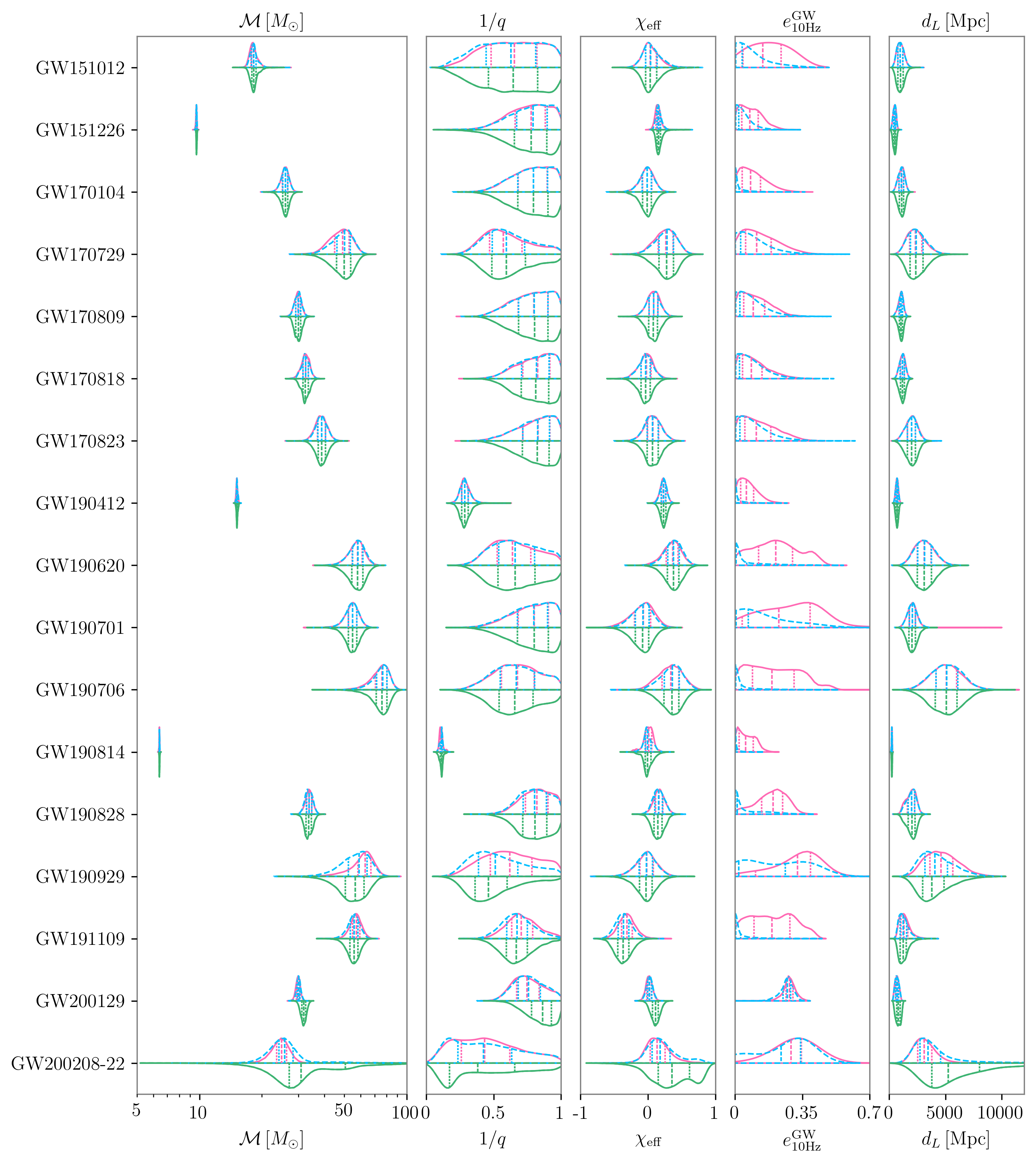}
    \caption{
    Marginal probability distributions for the chirp mass $\mathcal{M}$ in solar masses, inverse mass ratio $1/q$, effective spin $\chieff$, reference GW eccentricity $e^{\mathrm{GW}}$, and luminosity distance $d_L$ in Mpc for all analyzed events in this work, at a reference frequency of 10 Hz. 
    The upper half of each violin plot represents the marginal posterior distributions for \phTE using a uniform eccentricity prior (pink) and a logarithmic eccentricity prior (dashed blue), while the lower half shows those of the QC \phTHM model (green). Vertical lines indicate the $90\%$ credible intervals of each distribution.} 
    \label{fig:posteriors}
\end{figure*}
\section{Results}\label{sec:results}

Figure~\ref{fig:posteriors} shows the marginal posterior probability distributions for the 17 BBH GW events analyzed in this study. The upper half of each violin plot presents the results obtained with the eccentric \phTE model, while the lower half shows the results for the QC \phTHM model. 
For the eccentric model, we include the distributions obtained with both the uniform and log-uniform eccentricity priors. In cases where no eccentricity signatures are found, the QC parameters exhibit qualitatively consistent distributions for both the \phTHM and \phTE models, regardless of the prior used.

A key outcome of this study is to demonstrate the ability of the \phTE model to perform PE on large sets of events, including events with low total mass such as GW190814 \cite{GW190814}. This event, as found by the LVK Collaboration, corresponds to a compact binary system consisting of a $23.2^{+1.1}_{-1.0}M_{\odot}$ black hole and a $2.59^{+0.08}_{-0.09}M_{\odot}$ compact object, whose nature remains uncertain~\cite{GW190814}. 
We do not find evidence of eccentricity in GW190814. Our analysis yields a GW eccentricity of $e^{\rm{GW}}_{10\rm{Hz}}=0.06^{+0.08}_{-0.05}$ with a negative significance of $\log_{10}\mathcal{B}_{\mathrm{E/QC}}=-0.24^{+0.18}_{-0.18}$ for the uniform prior, and $e^{\rm{GW}}_{10\rm{Hz}}=0.00^{+0.04}_{-0.00}$ with $\log_{10}\mathcal{B}_{\mathrm{E/QC}}=0.04^{+0.18}_{-0.18}$ for the log-uniform prior, with negative values within errors. 
Notably, the model proves efficient in performing PE for this low-mass event, motivating a dedicated reanalysis of low-mass events in a companion paper~\cite{Lluc_NSBH_2025}, where we conduct the first full IMR study to search for eccentric features in these events.

Regarding the eccentricity posteriors, we find clear support for eccentricity in two events, GW200129, and GW200208\_22, regardless of the prior choice. Both events have shown to exhibit signatures of eccentricity in previous studies~\cite{Gupte:2024jfe, Romero-Shaw:2022xko}, and our analysis reveals distributions that agree across the different priors. These two events, however, warrant further investigation, and we present summarized results for both in Secs.~\ref{subsec:gw200129} and \ref{subsec:GW200208}.

\begin{table*}
\centering
\resizebox{\textwidth}{!}{
\begin{tabular}{@{}clcccccccccccc@{}}
\toprule
\textbf{Event} & \textbf{Model} & $M/M_\odot$ & $\mathcal{M}/M_\odot$ & $1/q$ & $\chi_\text{eff}$ & $\chi_\text{p}$ & $e^\text{GW}_{\text{10Hz}}$ & $l^\text{GW}_{\text{10Hz}}$ & $d_L$ & $\text{SNR}^{\text{N}}$ & $\log_{10}\mathcal{B}_{\mathrm{E(P)}/\mathrm{QC}}$ \\ 
\midrule
\multirow{6}{*}{\texttt{GW190701}}
& \phTHM  &$127.74^{+14.41}_{-13.52}$ & $54.77^{+6.56}_{-6.90}$ & $0.80^{+0.18}_{-0.30}$ & $-0.08^{+0.22}_{-0.28}$ & -- & -- & -- & $2034^{+733}_{-676}$ & $11.38^{+0.13}_{-0.25}$ & -- \\
& \phTE$^{\mathrm{Uni}}$  &$127.75^{+14.33}_{-13.93}$ & $54.89^{+6.35}_{-7.02}$ & $0.80^{+0.18}_{-0.29}$ & $-0.07^{+0.24}_{-0.33}$ & -- & $0.29^{+0.17}_{-0.28}$ & $3.41^{+2.54}_{-3.07}$ & $2054^{+700}_{-691}$ & $11.60^{+0.42}_{-0.39}$  & $0.23^{+0.13}_{-0.13}$ \\
& \phTE$^{\mathrm{Uni}}_{\texttt{nlive=}2000}$ & $126.97^{+15.26}_{-14.17}$ & $54.47^{+6.88}_{-7.41}$ & $0.80^{+0.18}_{-0.31}$ & $-0.09^{+0.26}_{-0.32}$ & -- & $0.29^{+0.17}_{-0.28}$ & $3.35^{+2.61}_{-3.06}$ & $2024^{+715}_{-693}$ & $11.56^{+0.39}_{-0.36}$ & $0.24^{+0.11}_{-0.11}$ \\
& \phTE$^{\mathrm{LogUni}}$  & $127.83^{+14.80}_{-13.83}$ & $54.85^{+6.67}_{-7.06}$ & $0.80^{+0.18}_{-0.29}$ & $-0.08^{+0.23}_{-0.29}$ & -- & $0.06^{+0.34}_{-0.06}$ & $3.21^{+2.77}_{-2.89}$ & $2055^{+731}_{-683}$ & $11.41^{+0.22}_{-0.28}$ & $0.08^{+0.12}_{-0.12}$ \\
& \phTE$^{\mathrm{LogUni}}_{\texttt{nlive=}2000}$ & $127.71^{+14.90}_{-14.14}$ & $54.81^{+6.73}_{-7.35}$ & $0.80^{+0.18}_{-0.30}$ & $-0.08^{+0.23}_{-0.29}$ & -- & $0.06^{+0.35}_{-0.06}$ & $3.22^{+2.76}_{-2.87}$ & $2040^{+724}_{-676}$ & $11.40^{+0.29}_{-0.27}$ & $0.09^{+0.10}_{-0.10}$  \\
& \phTPHM  & $130.58^{+17.82}_{-14.80}$ & $55.98^{+7.91}_{-7.68}$ & $0.80^{+0.18}_{-0.31}$ & $-0.06^{+0.24}_{-0.29}$ & $0.46^{+0.40}_{-0.35}$ & -- & -- & $2126^{+753}_{-711}$ & $11.37^{+0.15}_{-0.28}$ & $0.05^{+0.12}_{-0.12}$ \\
\midrule
\multirow{6}{*}{\texttt{GW190929}}
& \phTHM  &$141.64^{+23.26}_{-20.72}$ & $56.32^{+13.07}_{-13.67}$ & $0.46^{+0.41}_{-0.20}$ & $-0.03^{+0.23}_{-0.27}$ & -- & -- & -- & $3795^{+2916}_{-1675}$ & $9.96^{+0.40}_{-0.52}$  & -- &\\
& \phTE$^{\mathrm{Uni}}$  &$150.66^{+22.43}_{-21.91}$ & $62.99^{+10.09}_{-14.75}$ & $0.61^{+0.34}_{-0.30}$ & $-0.02^{+0.24}_{-0.27}$ & -- & $0.35^{+0.08}_{-0.32}$ & $3.64^{+2.24}_{-3.19}$ & $4566^{+2692}_{-2042}$ & $10.24^{+0.39}_{-0.53}$& $0.56^{+0.11}_{-0.11}$ \\
& \phTE$^{\mathrm{Uni}}_{\texttt{nlive=}2000}$ & $150.33^{+22.74}_{-21.26}$ & $63.00^{+10.16}_{-15.03}$ & $0.62^{+0.33}_{-0.31}$ & $-0.02^{+0.24}_{-0.28}$ & -- & $0.35^{+0.08}_{-0.32}$ & $3.57^{+2.35}_{-3.13}$ & $4602^{+2618}_{-2042}$ & $10.23^{+0.40}_{-0.52}$ &  $0.56^{+0.09}_{-0.09}$\\
& \phTE$^{\mathrm{LogUni}}$  &  $144.78^{+23.75}_{-21.91}$ & $59.03^{+12.31}_{-15.37}$ & $0.51^{+0.40}_{-0.24}$ & $-0.02^{+0.23}_{-0.27}$ & -- & $0.18^{+0.25}_{-0.17}$ & $3.43^{+2.54}_{-3.08}$ & $4048^{+2878}_{-1821}$ & $10.07^{+0.46}_{-0.57}$ & $0.19^{+0.10}_{-0.10}$ \\
& \phTE$^{\mathrm{LogUni}}_{\texttt{nlive=}2000}$ & $144.59^{+23.86}_{-21.98}$ & $58.81^{+12.54}_{-15.33}$ & $0.51^{+0.41}_{-0.25}$ & $-0.02^{+0.24}_{-0.28}$ & -- & $0.19^{+0.24}_{-0.18}$ & $3.32^{+2.64}_{-2.97}$ & $4060^{+2915}_{-1822}$ & $10.07^{+0.46}_{-0.57}$ &  $0.19^{+0.09}_{-0.09}$\\
& \phTPHM  & $146.93^{+41.35}_{-21.57}$ & $57.46^{+18.60}_{-14.45}$ & $0.44^{+0.42}_{-0.22}$ & $0.03^{+0.28}_{-0.26}$ & $0.39^{+0.43}_{-0.30}$ & -- & -- & $3716^{+3077}_{-1626}$ & $10.03^{+0.59}_{-0.58}$ & $0.09^{+0.11}_{-0.11}$ \\
\midrule
\multirow{3}{*}{\shortstack{\texttt{GW200129} \\ \texttt{GWOSC}}}
& \phTHM  & $73.54^{+3.52}_{-3.28}$ & $31.83^{+1.56}_{-1.56}$ & $0.86^{+0.12}_{-0.19}$ & $0.11^{+0.10}_{-0.11}$ & -- & -- & -- & $818^{+317}_{-347}$ & $26.27^{+0.13}_{-0.17}$ & -- \\
& \phTE$^{\mathrm{Uni}}$  & $69.81^{+2.51}_{-2.34}$ & $29.93^{+1.17}_{-1.23}$ & $0.75^{+0.21}_{-0.16}$ & $0.02^{+0.08}_{-0.08}$ & -- & $0.28^{+0.05}_{-0.06}$ & $2.77^{+3.10}_{-2.40}$ & $686^{+288}_{-272}$ & $26.94^{+0.13}_{-0.20}$  &  $5.14^{+0.15}_{-0.15}$\\
& \phTE$^{\mathrm{LogUni}}$  & $70.18^{+2.33}_{-2.24}$ & $30.09^{+1.17}_{-1.10}$ & $0.76^{+0.19}_{-0.17}$ & $0.02^{+0.08}_{-0.08}$ & -- & $0.27^{+0.05}_{-0.07}$ & $2.60^{+3.35}_{-2.29}$ & $689^{+297}_{-257}$ & $26.92^{+0.15}_{-0.20}$ & $4.19^{+0.15}_{-0.15}$  \\
& \phTHM$_{\texttt{nlive=}2000}$(H1)  & $74.59^{+5.79}_{-5.48}$ & $32.17^{+2.52}_{-2.61}$ & $0.83^{+0.15}_{-0.23}$ & $0.12^{+0.16}_{-0.17}$ & -- & -- & -- & $1168^{+586}_{-585}$ & $14.73^{+0.11}_{-0.20}$ & -- \\
& \phTE$^{\mathrm{Uni}}_{\texttt{nlive=}2000}$(H1)  & $72.90^{+5.76}_{-4.54}$ & $31.41^{+2.57}_{-2.20}$ & $0.82^{+0.16}_{-0.23}$ & $0.09^{+0.16}_{-0.15}$ & -- & $0.18^{+0.14}_{-0.16}$ & $3.08^{+2.88}_{-2.76}$ & $1166^{+535}_{-578}$ & $14.83^{+0.19}_{-0.26}$  & $-0.06^{+0.08}_{-0.08}$ \\
& \phTE$^{\mathrm{LogUni}}_{\texttt{nlive=}2000}$(H1) & $74.37^{+5.72}_{-5.14}$ & $32.07^{+2.51}_{-2.42}$ & $0.83^{+0.15}_{-0.22}$ & $0.12^{+0.16}_{-0.16}$ & -- & $0.01^{+0.21}_{-0.01}$ & $3.15^{+2.82}_{-2.81}$ & $1181^{+563}_{-596}$ & $14.75^{+0.15}_{-0.21}$ & $0.02^{+0.08}_{-0.08}$  \\
\cdashline{1-13}
\multirow{5}{*}{\shortstack{\texttt{GW200129} \\ \texttt{gw\_subtract}}}
& \phTHM  & $73.05^{+3.53}_{-3.18}$ & $31.49^{+1.60}_{-1.53}$ & $0.81^{+0.17}_{-0.18}$ & $0.09^{+0.10}_{-0.11}$ & -- & -- & -- & $894^{+328}_{-367}$ & $25.93^{+0.12}_{-0.17}$  & -- \\
& \phTE$^{\mathrm{Uni}}$  & $70.42^{+2.36}_{-1.94}$ & $30.08^{+1.12}_{-0.96}$ & $0.71^{+0.18}_{-0.13}$ & $0.02^{+0.08}_{-0.07}$ & -- & $0.26^{+0.04}_{-0.07}$ & $2.72^{+3.29}_{-2.42}$ & $775^{+290}_{-310}$ & $26.49^{+0.14}_{-0.20}$ & $4.00^{+0.15}_{-0.15}$ \\
& \phTE$^{\mathrm{LogUni}}$  & $70.64^{+2.30}_{-2.02}$ & $30.20^{+1.11}_{-0.98}$ & $0.72^{+0.18}_{-0.14}$ & $0.03^{+0.08}_{-0.07}$ & -- & $0.26^{+0.04}_{-0.07}$ & $2.79^{+3.20}_{-2.53}$ & $773^{+290}_{-311}$ & $26.49^{+0.15}_{-0.20}$ &  $3.35^{+0.15}_{-0.15}$\\
& \phTPHM  &$73.28^{+4.04}_{-3.36}$ & $31.56^{+1.71}_{-1.61}$ & $0.79^{+0.18}_{-0.18}$ & $0.08^{+0.11}_{-0.11}$ & $0.42^{+0.39}_{-0.30}$ & -- & -- & $896^{+323}_{-333}$ & $25.99^{+0.17}_{-0.19}$  & $0.11^{+0.14}_{-0.14}$ \\
& \NRSur  & $73.31^{+3.38}_{-2.79}$ & $30.18^{+2.01}_{-1.71}$ & $0.54^{+0.38}_{-0.13}$ & $0.00^{+0.12}_{-0.12}$ & $0.81^{+0.15}_{-0.54}$ & -- & -- & $1121^{+193}_{-287}$ & $26.20^{+0.16}_{-0.23}$ & $2.12^{+0.14}_{-0.14}$ \\
\cdashline{1-13}
\multirow{5}{*}{\shortstack{\texttt{GW200129} \\ \texttt{BayesA}}}
& \phTHM  & $76.65^{+3.56}_{-3.26}$ & $33.19^{+1.55}_{-1.55}$ & $0.86^{+0.13}_{-0.17}$ & $0.22^{+0.09}_{-0.10}$ & -- & -- & -- & $967^{+325}_{-405}$ & $26.71^{+0.11}_{-0.15}$ & --  &\\
& \phTE$^{\mathrm{Uni}}$  & $74.32^{+2.66}_{-2.22}$ & $32.05^{+1.24}_{-1.03}$ & $0.81^{+0.16}_{-0.15}$ & $0.17^{+0.08}_{-0.07}$ & -- & $0.21^{+0.05}_{-0.08}$ & $2.93^{+3.13}_{-2.70}$ & $854^{+332}_{-349}$ & $27.04^{+0.14}_{-0.19}$ & $1.86^{+0.15}_{-0.15}$ \\
& \phTE$^{\mathrm{LogUni}}$  & $74.64^{+2.92}_{-2.27}$ & $32.22^{+1.35}_{-0.98}$ & $0.82^{+0.15}_{-0.14}$ & $0.18^{+0.08}_{-0.07}$ & -- &  $0.19^{+0.06}_{-0.14}$ & $3.28^{+2.69}_{-2.97}$ & $858^{+336}_{-337}$ & $27.03^{+0.16}_{-0.26}$ &  $1.30^{+0.15}_{-0.15}$\\
& \phTPHM  & $77.54^{+3.66}_{-3.75}$ & $33.52^{+1.48}_{-1.66}$ & $0.84^{+0.14}_{-0.16}$ & $0.23^{+0.09}_{-0.11}$ & $0.48^{+0.34}_{-0.29}$ & -- & -- & $1061^{+268}_{-398}$ & $26.76^{+0.12}_{-0.16}$ & $0.11^{+0.15}_{-0.15}$ \\
& \NRSur  & $77.26^{+3.70}_{-4.32}$ & $33.13^{+1.74}_{-1.99}$ & $0.77^{+0.20}_{-0.21}$ & $0.19^{+0.10}_{-0.12}$ & $0.58^{+0.33}_{-0.36}$ & -- & -- & $1170^{+259}_{-440}$ & $26.84^{+0.12}_{-0.17}$ &  $1.17^{+0.15}_{-0.15}$ \\
\cdashline{1-13}
\multirow{5}{*}{\shortstack{\texttt{GW200129} \\ \texttt{BayesB}}}
& \phTHM  & $75.48^{+3.56}_{-3.40}$ & $32.63^{+1.59}_{-1.54}$ & $0.85^{+0.13}_{-0.18}$ & $0.18^{+0.10}_{-0.10}$ & -- & -- & -- & $968^{+325}_{-400}$ & $26.30^{+0.12}_{-0.15}$  & -- \\
& \phTE$^{\mathrm{Uni}}$  & $72.69^{+2.52}_{-2.29}$ & $31.26^{+1.15}_{-1.09}$ & $0.77^{+0.18}_{-0.15}$ & $0.11^{+0.08}_{-0.08}$ & -- & $0.24^{+0.04}_{-0.07}$ & $2.95^{+3.10}_{-2.72}$ & $806^{+324}_{-325}$ & $26.75^{+0.15}_{-0.19}$ & $2.90^{+0.15}_{-0.15}$  \\
& \phTE$^{\mathrm{LogUni}}$  & $73.11^{+2.70}_{-2.30}$ & $31.46^{+1.26}_{-1.09}$ & $0.78^{+0.17}_{-0.14}$ & $0.12^{+0.08}_{-0.08}$ & -- &  $0.22^{+0.05}_{-0.08}$ & $3.00^{+3.05}_{-2.79}$ & $820^{+341}_{-351}$ & $26.73^{+0.15}_{-0.21}$ &  $2.13^{+0.15}_{-0.15}$\\
& \phTPHM  &$75.95^{+3.63}_{-3.56}$ & $32.80^{+1.55}_{-1.67}$ & $0.82^{+0.16}_{-0.16}$ & $0.17^{+0.10}_{-0.11}$ & $0.46^{+0.36}_{-0.30}$ & - & - & $1028^{+286}_{-378}$ & $26.35^{+0.13}_{-0.16}$ & $0.04^{+0.14}_{-0.14}$ \\
& \NRSur  & $75.15^{+4.62}_{-3.27}$ & $31.69^{+2.47}_{-2.05}$ & $0.67^{+0.29}_{-0.20}$ & $0.11^{+0.13}_{-0.13}$ & $0.75^{+0.21}_{-0.47}$ & -- & -- & $1182^{+220}_{-431}$ & $26.51^{+0.15}_{-0.18}$ & $1.70^{+0.14}_{-0.14}$ \\
\cdashline{1-13}
\multirow{2}{*}{\shortstack{\texttt{GW200129} \\ \texttt{BayesC}}}
& \phTHM  &$75.00^{+3.62}_{-3.24}$ & $32.47^{+1.55}_{-1.52}$ & $0.86^{+0.13}_{-0.18}$ & $0.16^{+0.10}_{-0.10}$ & -- & -- & -- & $987^{+324}_{-415}$ & $25.59^{+0.12}_{-0.16}$ & --  &\\
& \phTE$^{\mathrm{Uni}}$  & $72.55^{+2.61}_{-2.44}$ & $31.23^{+1.22}_{-1.18}$ & $0.79^{+0.17}_{-0.16}$ & $0.10^{+0.08}_{-0.08}$ & -- & $0.24^{+0.04}_{-0.08}$ & $2.78^{+3.27}_{-2.56}$  & $831^{+347}_{-348}$ & $26.05^{+0.15}_{-0.20}$ &  $2.82^{+0.15}_{-0.15}$ \\
\midrule
\multirow{7}{*}{\texttt{GW200208\_22}}
& \phTHM  &$90.39^{+108.92}_{-29.85}$ & $30.32^{+48.65}_{-8.45}$ & $0.39^{+0.53}_{-0.30}$ & $0.37^{+0.45}_{-0.45}$ & -- & -- & -- & $5066^{+9043}_{-2711}$ & $7.22^{+0.56}_{-1.89}$ & --  \\
& \phTHM$_{\texttt{nlive=}2000}$  &$95.10^{+153.18}_{-34.29}$ & $30.91^{+62.42}_{-8.73}$ & $0.38^{+0.54}_{-0.29}$ & $0.36^{+0.46}_{-0.46}$ & -- & -- & -- & $5247^{+10830}_{-2852}$ & $7.19^{+0.58}_{-3.49}$ &   \\
& \phTE$^{\mathrm{Uni}}$  & $65.35^{+45.54}_{-7.84}$ & $24.95^{+4.62}_{-3.57}$ & $0.43^{+0.46}_{-0.31}$ & $0.13^{+0.33}_{-0.23}$ & -- & $0.33^{+0.05}_{-0.16}$ & $2.99^{+2.94}_{-2.69}$ & $3101^{+2176}_{-1264}$ & $8.37^{+0.53}_{-0.98}$ &  $0.96^{+0.12}_{-0.12}$  \\
& \phTE$^{\mathrm{Uni}}_{\texttt{nlive=}2000}$ & $65.75^{+52.42}_{-8.39}$ & $24.73^{+6.41}_{-4.05}$ & $0.39^{+0.48}_{-0.27}$ & $0.15^{+0.31}_{-0.24}$ & -- & $0.34^{+0.08}_{-0.18}$ & $3.07^{+2.87}_{-2.74}$ & $3021^{+2662}_{-1260}$ & $8.44^{+0.64}_{-1.15}$ & $1.14^{+0.08}_{-0.08}$  \\
& \phTE$^{\mathrm{LogUni}}$ & $68.43^{+112.66}_{-10.34}$ & $25.82^{+39.26}_{-4.19}$ & $0.42^{+0.48}_{-0.31}$ & $0.18^{+0.51}_{-0.28}$ & -- & $0.31^{+0.09}_{-0.30}$ & $3.02^{+2.95}_{-2.72}$ & $3466^{+7332}_{-1504}$ & $8.09^{+0.76}_{-2.17}$ & $0.09^{+0.11}_{-0.11}$ \\
& \phTE$^{\mathrm{LogUni}}_{\texttt{nlive=}2000}$ & $68.54^{+114.84}_{-10.36}$ & $25.95^{+40.04}_{-4.35}$ & $0.43^{+0.48}_{-0.32}$ & $0.19^{+0.52}_{-0.29}$ & $0.00^{+0.00}_{-0.00}$ & $0.29^{+0.09}_{-0.28}$ & $3.07^{+2.90}_{-2.78}$ & $3482^{+7371}_{-1539}$ & $8.06^{+0.77}_{-2.17}$  &$0.49^{+0.08}_{-0.08}$  \\
& \phTPHM  & $99.61^{+136.58}_{-38.39}$ & $33.15^{+57.49}_{-9.57}$ & $0.42^{+0.50}_{-0.32}$ & $0.36^{+0.47}_{-0.48}$ & $0.41^{+0.42}_{-0.31}$ & - & - & $5308^{+10430}_{-2855}$ & $7.22^{+0.68}_{-2.38}$ & $-0.26^{+0.10}_{-0.10}$  \\
&\phXPHM  & $173.89^{+100.26}_{-106.75}$ & $45.51^{+31.26}_{-19.64}$ & $0.16^{+0.63}_{-0.10}$ & $0.67^{+0.21}_{-0.62}$ & $0.43^{+0.37}_{-0.29}$ & - & - & $5816^{+7900}_{-2947}$ & $7.33^{+1.15}_{-2.04}$ & --  &\\

\bottomrule
\end{tabular}
}
\caption{Median values and 90\% credible intervals of the posterior distributions for the analyzed GW events (indicated in each row). The parameters displayed are the total mass $M$ and chirp mass $\mathcal{M}$ in solar masses (both in the detector's frame), the inverse mass ratio $1/q$, the effective spin parameter $\chi_{\mathrm{eff}}$, the reference GW eccentricity, $e^{\rm GW}_{\text{10Hz}}$, and GW mean anomaly $l^{\rm GW}_{\text{10Hz}}$, the luminosity distance $d_L$, and the network matched-filtered SNR, $\text{SNR}^{\mathrm{N}}$. The last column shows the log-10 Bayes factor between the eccentric (E) and the QC aligned-spin (QC) hypothesis $\log_{10}\mathcal{B}_{\mathrm{E}/\mathrm{QC}}$, or between the the QC precessing-spin (P) and QC hypothesis $\log_{10}\mathcal{B}_{\mathrm{P}/\mathrm{QC}}$. The spins and eccentric parameters are given at the reference frequency of 10 Hz. 
For GW200129, we present results obtained using different datasets: first, the non-deglitched publicly released \texttt{GWOSC} data~\cite{gwosc3}, followed by the \texttt{gw\_subtract} data, and finally, the \texttt{BayesWave} technique, which provides three distinct draws: \texttt{BayesA}, \texttt{BayesB}, and \texttt{BayesC}.
For GW200208\_22, we include the official LVK samples using the \phXPHM model from GWTC-3~\cite{gwtc3}. }
\label{tab:GW_pes}
\end{table*}

\begin{figure*}
    \centering
    \includegraphics[width=1\linewidth]{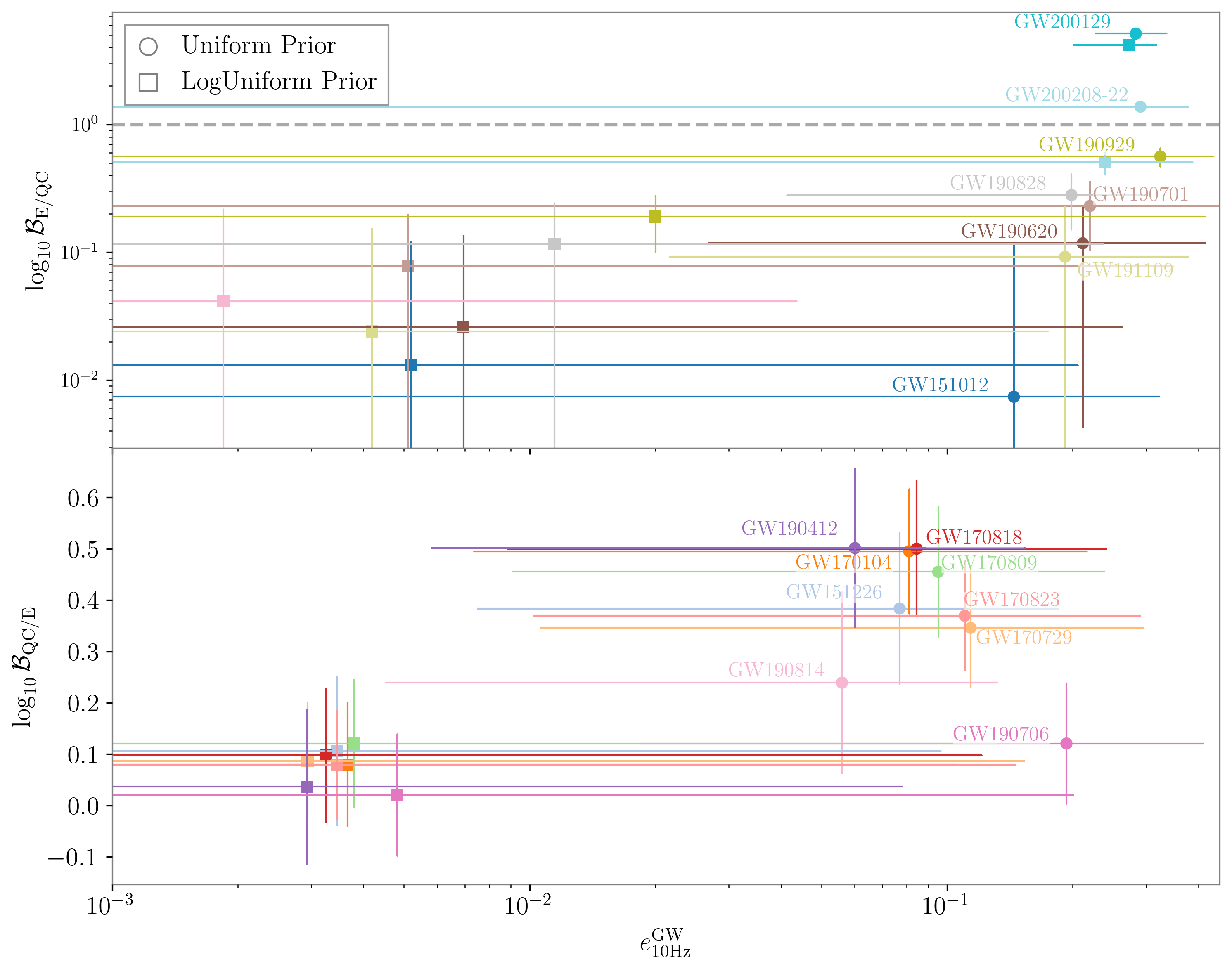}
    \caption{
    Log-10 Bayes factors comparing the eccentric (E) and aligned quasi-circular (QC) hypotheses for each analyzed GW event, plotted against the mean inferred GW eccentricity at the reference frequency, $e^{\mathrm{GW}}_{10\mathrm{Hz}}$.
    The top panel displays the factor for cases that favor the eccentric hypothesis ($\log_{10}\mathcal{B}_{\mathrm{E/QC}}\geq0$), while the bottom panel shows the events that favor the QC hypothesis ($\log_{10}\mathcal{B}_{\mathrm{QC/E}}\geq0$).
    In the top panel, we include a dashed gray line at $\log_{10}\mathcal{B}_{\mathrm{E/QC}}=0$ to indicate the threshold for support in favor of the eccentric hypothesis.
    Circles denote results obtained using a uniform eccentricity prior, whereas squares represent those using a logarithmic prior. Horizontal lines indicate the 90\% credible intervals for the GW eccentricity posteriors. Different colors distinguish the GW events, with labels placed next to their corresponding uniform prior points.}
    \label{fig:bayes}
\end{figure*}

We compute the log-10 Bayes factors between the eccentric and the QC hypotheses (both aligned and precessing spins), as defined in Eq.~\eqref{eq:bayes}, and display them as a function of the measured eccentricity in Fig.~\ref{fig:bayes}. 
For GW200129, we observe a high value of the Bayes factor in favour of the eccentric hypothesis, which is also supported when using a log-uniform prior. 
In contrast, for GW200208\_22, we find a $\log_{10}\mathcal{B}_{\mathrm{E/QC}}$ greater than 1 only when using the uniform prior. Detailed studies on both events are provided in Secs.~\ref{subsec:gw200129} and \ref{subsec:GW200208}, respectively.
There are two additional events of particular interest. The first is GW190701, for which Ref.~\cite{Gupte:2024jfe} finds support for eccentricity, while our analysis notably reduces this support when using a log-uniform prior. The second event is GW190929, where we observe some support for non-zero eccentricity. Nevertheless, for both cases the log-uniform prior shifts the eccentricity values lower. Still, the eccentric hypothesis cannot be completely ruled out, as residual support for eccentricity persists even when using finer sampler settings.
These two events are studied in more detail in Sec.~\ref{subsec:highmass}.

Furthermore, the posteriors for these four events differ significantly between the eccentric and QC waveform model. In particular, the last two rows of Fig.~\ref{fig:posteriors} show notable discrepancies in the chirp mass and the effective spin parameters between the QC and eccentric model results. 
A more detailed discussion of these discrepancies is provided in the following sections.

\subsection{GW200129}\label{subsec:gw200129}

GW200129 is an event of particular interest due to various imprints suggesting a possible dynamical formation channel. 
Additionally, the binary's low mass --allowing for several inspiral cycles within the detector's frequency band-- and its relatively high SNR ($\sim 26$, see Tab.~\ref{tab:GW_pes} for details) make it well-suited for characterizing these effects.
Previous studies have identified signatures of spin precession~\cite{Hannam_2022}, evidence of a significant kick velocity~\cite{Varma_2022}, and indications of orbital eccentricity~\cite{Gupte:2024jfe, Romero-Shaw:2022xko}, as well as false violations of general relativity due to waveform systematics~\cite{Maggio_2023}. 
The GWTC-3 catalog~\cite{gwtc3} reports an effective-spin parameter of $\chieff=0.11^{+0.11}_{-0.16}$, a total mass of $M=63.3^{+4.5}_{-3.4}M_{\odot}$, a luminosity distance of $d_L=890^{+260}_{-370}$ Mpc, and a network matched-filter signal-to-noise ratio of $\rm SNR=26.8^{+0.2}_{-0.2}$.
However, this event also presents a glitch in LIGO Livingston data within the 20–50 Hz frequency range, and different glitch mitigation techniques have been shown to significantly affect the support for spin precession~\cite{Payne_2022,Macas_2024}.
\begin{figure*}
    \centering
    GW200129
    \includegraphics[width=\linewidth]{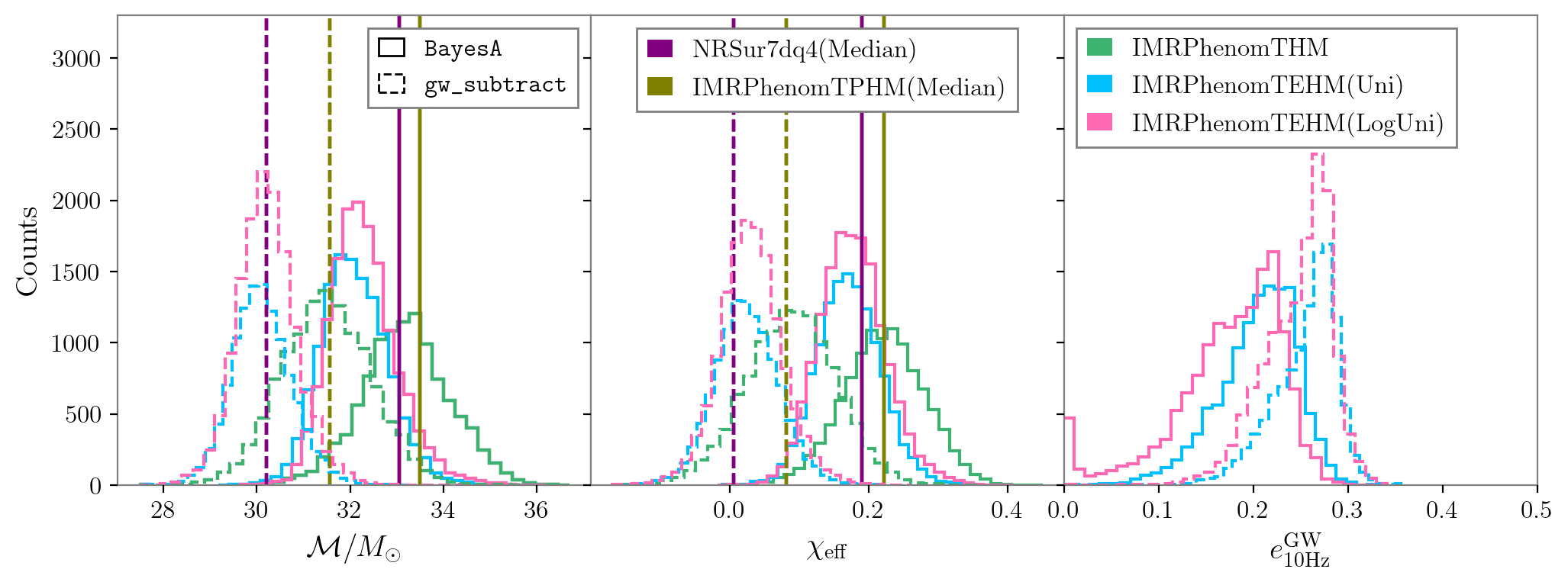}
    \caption{Posterior distributions for the chirp mass, effective spin, and GW eccentricity at a reference frequency of 10 Hz for GW200129. Results are shown for the \texttt{gw\_subtract} (\textit{dashed-lines}) and \texttt{BayesA} (\textit{continuous-lines}) glitch mitigation techniques. The distributions are provided for both the QC \phTHM (\textit{green}) and eccentric \phTE models, using uniform (Uni, \textit{blue}) and log-uniform (LogUni, \textit{pink}) priors. The median values for the \phTPHM (\textit{dark green}) and \NRSur (\textit{purple}) runs are included for comparison to highlight the impact of precessing features on the posterior distributions for both datasets.}
    \label{fig:GW200129}
\end{figure*}

To further investigate the nature of GW200129, we present the first eccentric analysis using standard PE techniques typically applied to QC binaries~\cite{gwtc3}. Additionally, we extensively investigate various glitch mitigation techniques and their impact on eccentricity recovery, employing both a uniform eccentricity prior and a log-uniform prior.
Specifically, we consider data obtained directly from GWOSC for all three detectors, as well as the glitch-mitigated data from Ref.~\cite{Payne_2022}, where the \texttt{gw\_subtract} mitigation technique and \texttt{BayesWave} were applied. For the latter, we use three different glitch draws, denoted as \texttt{BayesA}, \texttt{BayesB}, and \texttt{BayesC}. Due to inconsistencies identified in the Virgo detector, as detailed in Ref.~\cite{Payne_2022}, we restrict our deglitched analyses to the LIGO detectors.
Additionally, we analyze this event using the QC precessing-spin model of the \phT family, \phTPHM \cite{Estelles:2021gvs}, as well as the \NRSur model~\cite{Varma:2019vhw}.

Our findings are summarized in Tab.~\ref{tab:GW_pes} and are largely consistent with those reported in Ref.~\cite{Gupte:2024jfe}. 
We observe large variations in the measurement of parameters depending on the data used for the analysis. 
Specifically, the original data containing the glitch (labeled as \texttt{GWOSC}) favors the presence of orbital eccentricity over the QC hypothesis.  
The Bayes factors shown in Tab.~\ref{tab:GW_pes} strongly support, when using non-precessing models, the eccentric hypotheses over the QC scenario, with $\log_{10}\mathcal{B}_{\mathrm{E/QC}}$=$5.14^{+0.15}_{-0.15}$ and $\log_{10}\mathcal{B}_{\mathrm{E/QC}}$=$4.19^{+0.15}_{-0.15}$ when using a uniform and log-uniform prior in eccentricity, respectively. 
However, different glitch mitigation techniques introduce significant variations in these factors, as previously observed with precessing QC models. For this reason, we do not perform precessing QC runs on these non-deglitched data, as the noise artifact has been shown to limit the conclusions of the analysis~\cite{Hannam_2022, Payne_2022}.

The results obtained using \texttt{gw\_subtract} are shown in Tab.~\ref{tab:GW_pes}. For these data, we perform PE runs using all models, \phTHM, \phTPHM, \NRSur, and \phTE, both using a uniform and log-uniform prior in eccentricity. 
Figure~\ref{fig:GW200129} shows the posterior distributions for the chirp mass, effective spin, and eccentricity evaluated at a reference frequency of 10 Hz. 
Results for \phTE include both uniform (denoted as Uni) and log-uniform (LogUni) eccentricity priors. Furthermore, the median values obtained from the QC precessing-spin \phTPHM\ and \NRSur runs are indicated in  Fig.~\ref{fig:GW200129}. 
These results closely align with the results obtained with the \texttt{GWOSC} dataset, exhibiting strong support for a GW eccentricity of $e^{\rm{GW}}_{10\mathrm{Hz}}=0.26^{+0.04}_{-0.07}$ consistent between the uniform and log-uniform prior, with Bayes factors of $4.00^{+0.15}_{-0.15}$ and $3.35^{+0.15}_{-0.15}$, respectively. 
These values strongly suggest a preference for the eccentric hypothesis, though slightly lower than that obtained with the non-deglitched data. 
Notably, the QC precessing-spin hypothesis also remains favored over the aligned-spin QC scenario. As already reported in Ref.~\cite{Hannam_2022}, the \phTPHM model shows minor support for spin precession, and we find a Bayes factor of $\log_{10}\mathcal{B}_{\mathrm{P/QC}}$=$0.11^{+0.14}_{-0.14}$, while \NRSur is notably preferred over the QC aligned-spin hypothesis, with a Bayes factor of $\log_{10}\mathcal{B}_{\mathrm{P/QC}}$=$2.12^{+0.14}_{-0.14}$. However, these values remain significantly smaller than those obtained with the eccentric \phTE model, regardless of the prior choice. 
An interesting feature observable in Fig.~\ref{fig:GW200129} is that the median values of the recovered chirp mass and effective spin using \NRSur for \texttt{gw\_subtract} data (dashed lines in the plot) agree much better with the distributions obtained using \phTE rather than \phTHM. However, this is not the case for \phTPHM, which recovers values more consistent with those of the aligned-spin run (shown in light green in the plots). 
We thus find that the eccentric model posteriors agree better with the precessing surrogate, indicating a preference for dynamical effects in this event over the aligned-spin QC hypothesis, and highlighting the degeneracy between eccentricity and precession. For studies of similar degeneracies in the context of other events, see e.g. Refs.~\cite{PhysRevLett.126.201101,Romero-Shaw:2022fbf}.
A more detailed study would require the inclusion of spin-precessing effects in \phTE, which we leave for future work.

To visually assess the impact of incorporating eccentricity and to compare it with the precessing scenario, we show in the top panel of Fig.~\ref{fig:waveforms} the whitened data using the \texttt{gw\_subtract} data, overlaid with the reconstructed maximum likelihood waveforms from the two eccentric runs with \phTE using uniform and log-uniform eccentricity priors, the aligned-spin QC run with \phTHM, and the \NRSur waveform. Notably, the two eccentric reconstructions show excellent agreement with each other, while both differ from the QC models.

Recent studies on GW200129 have investigated the use of the \texttt{BayesWave} mitigation technique~\cite{Payne_2022,Macas_2024}. 
The analyses in GWTC-3~\cite{gwtc3} use a single draw of the glitch mitigation model, while a more detailed study on GW200129~\cite{Payne_2022} marginalizes over multiple draws. In this study, we employ the three glitch realizations provided in their public release~\cite{gwosc3}, labeled \texttt{BayesA}, \texttt{BayesB}, and \texttt{BayesC}. 
We perform parameter estimation with the three draws using the \phTHM and \phTE models with a uniform prior on eccentricity. Due to the high similarity between \texttt{BayesB} and \texttt{BayesC} (see Tab.~\ref{tab:GW_pes}), we exclude \texttt{BayesC} from further runs with the precessing-spin models and log-uniform prior. 
A summary of all performed runs is given in Tab.~\ref{tab:GW_pes}, and the results for \texttt{BayesA} --the run which exhibits the least support for dynamical features (both precession and eccentricity)-- are shown in Fig.~\ref{fig:GW200129}, along with those for \texttt{gw\_subtract} introduced above.
We find that the eccentric hypothesis is preferred over the precessing one for all the \texttt{BayesWave} draws, although this preference is not as strong as for the \texttt{gw\_subtract} data. While different mitigation techniques significantly impact the support for eccentricity, both in terms of recovered values and Bayes factors (see Tab.~\ref{tab:GW_pes}), the preference for the eccentric hypothesis over the precessing QC scenario persists overall.
The weakest support for eccentricity comes from the \texttt{BayesA} draw with a log-uniform prior, where the inferred eccentricity is
$e^{\mathrm{GW}}_{10\rm{Hz}}=0.19^{+0.06}_{-0.14}$, and the log-10 Bayes factor is $\log_{10}\mathcal{B}_{\mathrm{E/QC}}=1.30^{+0.15}_{-0.15}$. This data also shows the lowest preference for the precessing hypothesis in the \NRSur model, with a Bayes factor of $\log_{10}\mathcal{B}_{\mathrm{P/QC}}=1.17^{+0.15}_{-0.15}$, which, as noted before, remains lower than the support for eccentricity.

We also revisit the possibility of this event being QC spin-precessing, and we find consistently lower support for this hypothesis compared to the eccentric scenario, regardless of the model, prior, or data used, as shown in Tab.~\ref{tab:GW_pes}. Specifically, the preference for the eccentric hypothesis over the precessing one ranges from $\log_{10}\mathcal{B}_{\mathrm{E/P}}\in[0.13,1.88]$, as can be directly inferred from the table.

The variability in the evidence for eccentricity across different datasets suggests residual systematics in the glitch subtraction may still be influencing the analysis. 
Studies performing independent analyses on the LIGO detectors show that the support for eccentricity is much reduced when using only the LIGO Hanford detector, which is free from glitches. This is consistent with the lack of SNR in LIGO Hanford, due to the particular antenna patterns of this event~\cite{Gupte:2024jfe}.
To further investigate this, we performed three additional runs using only the Hanford data --listed in Tab.~\ref{tab:GW_pes} under ``H1''-- and found results consistent with those obtained using \texttt{BayesA} data, although with broader and less informative posteriors. Notably, the recovered Bayes factors are inconclusive, with values consistent with zero, in line with the substantially lower SNR (~15) compared to that obtained when including Livingston data (~27).
Nonetheless, our study consistently favors the eccentric hypothesis over the QC scenario, both for aligned and precessing spins.
Our results, together with previous studies~\cite{Hannam_2022,Gupte:2024jfe}, indicate a preference for dynamical formation channels, supported by spin precession or orbital eccentricity. However, distinguishing between eccentricity and precession or the combination of them requires generic spin eccentric waveform models. We plan to extend the \phTE waveform model to precessing-spin binaries in future work.

\subsection{GW200208\_22}\label{subsec:GW200208}
\begin{figure*}
    \centering
    \includegraphics[width=0.33\linewidth]{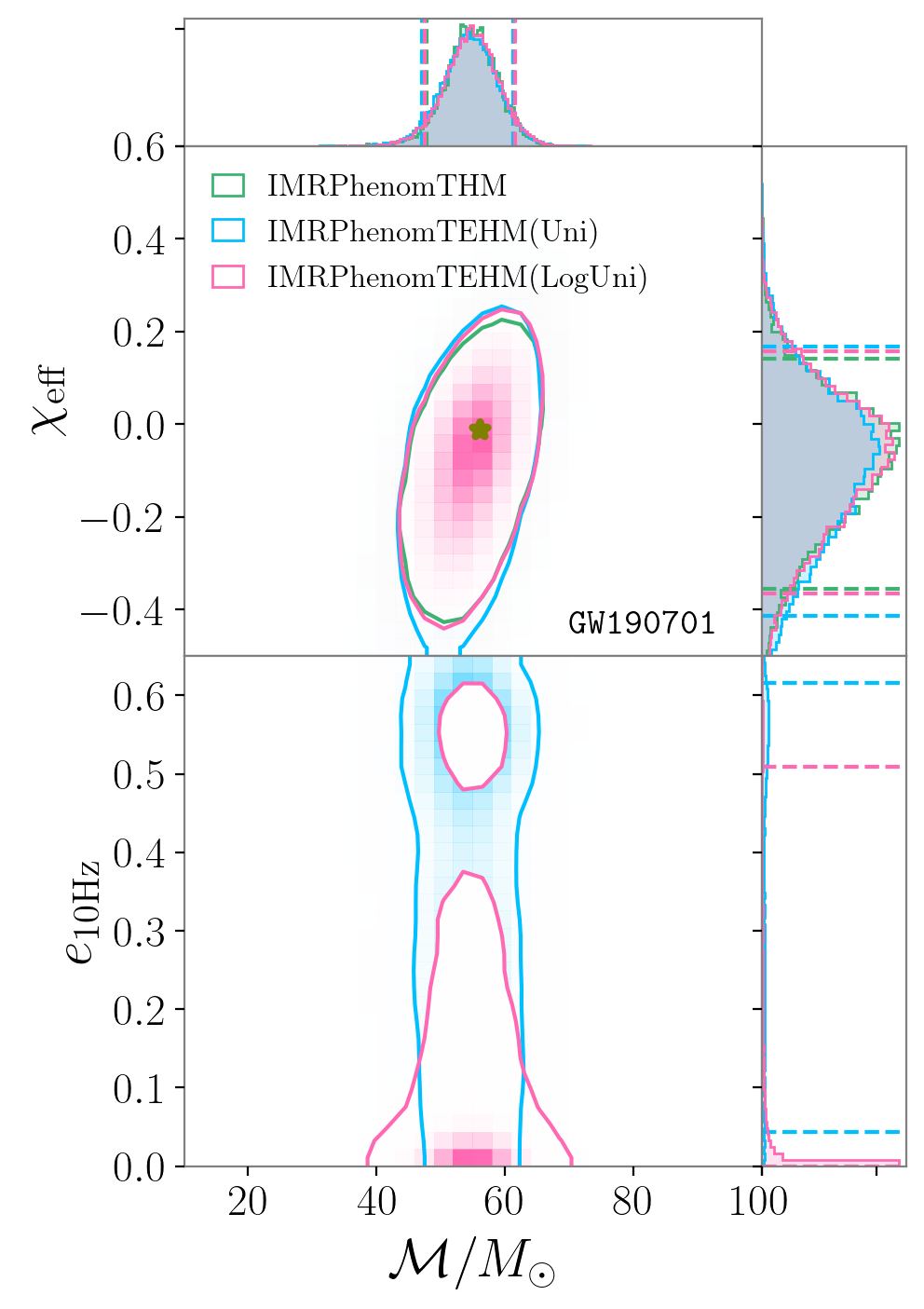}
    \includegraphics[width=0.33\linewidth]{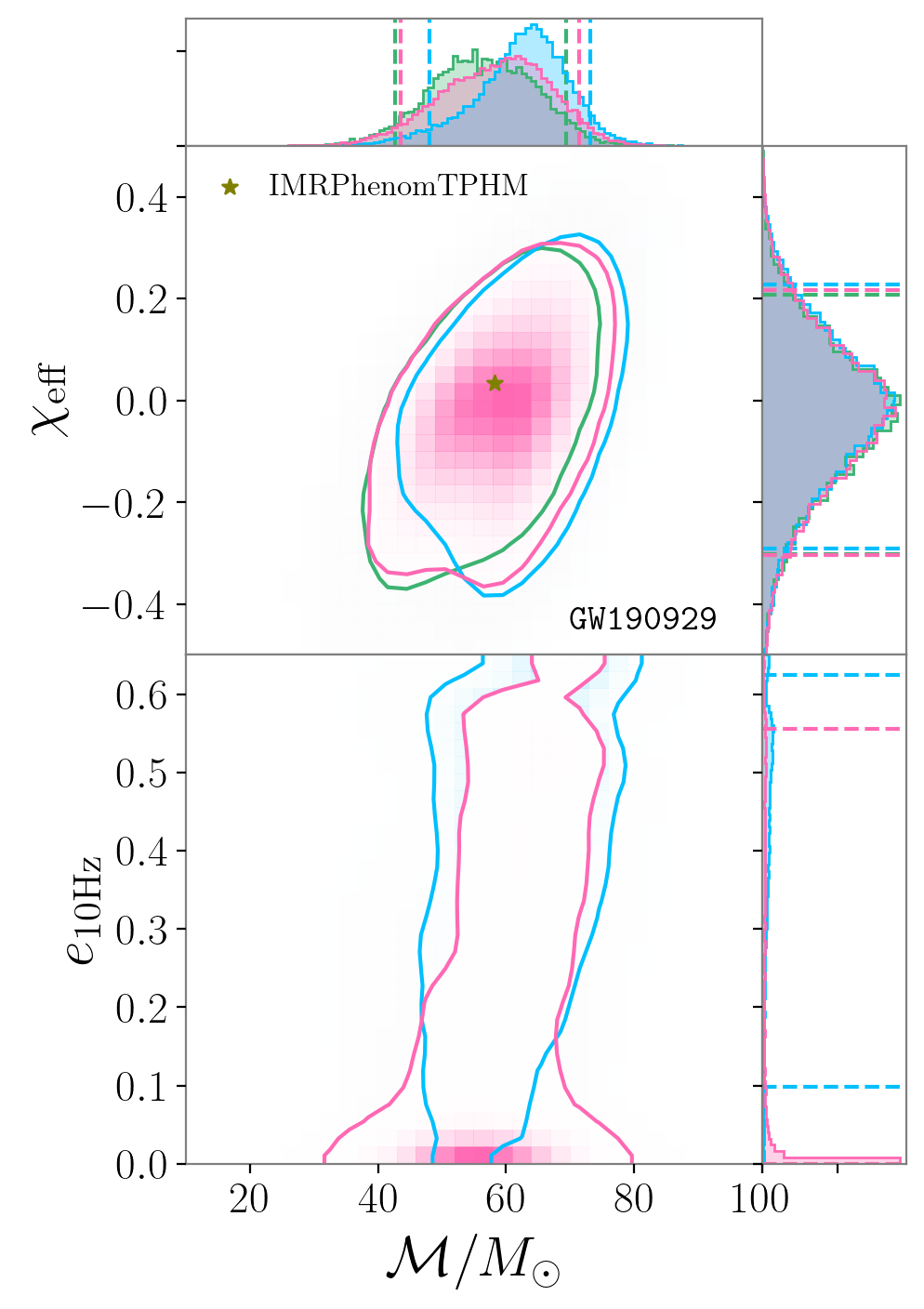}
    \includegraphics[width=0.33\linewidth]{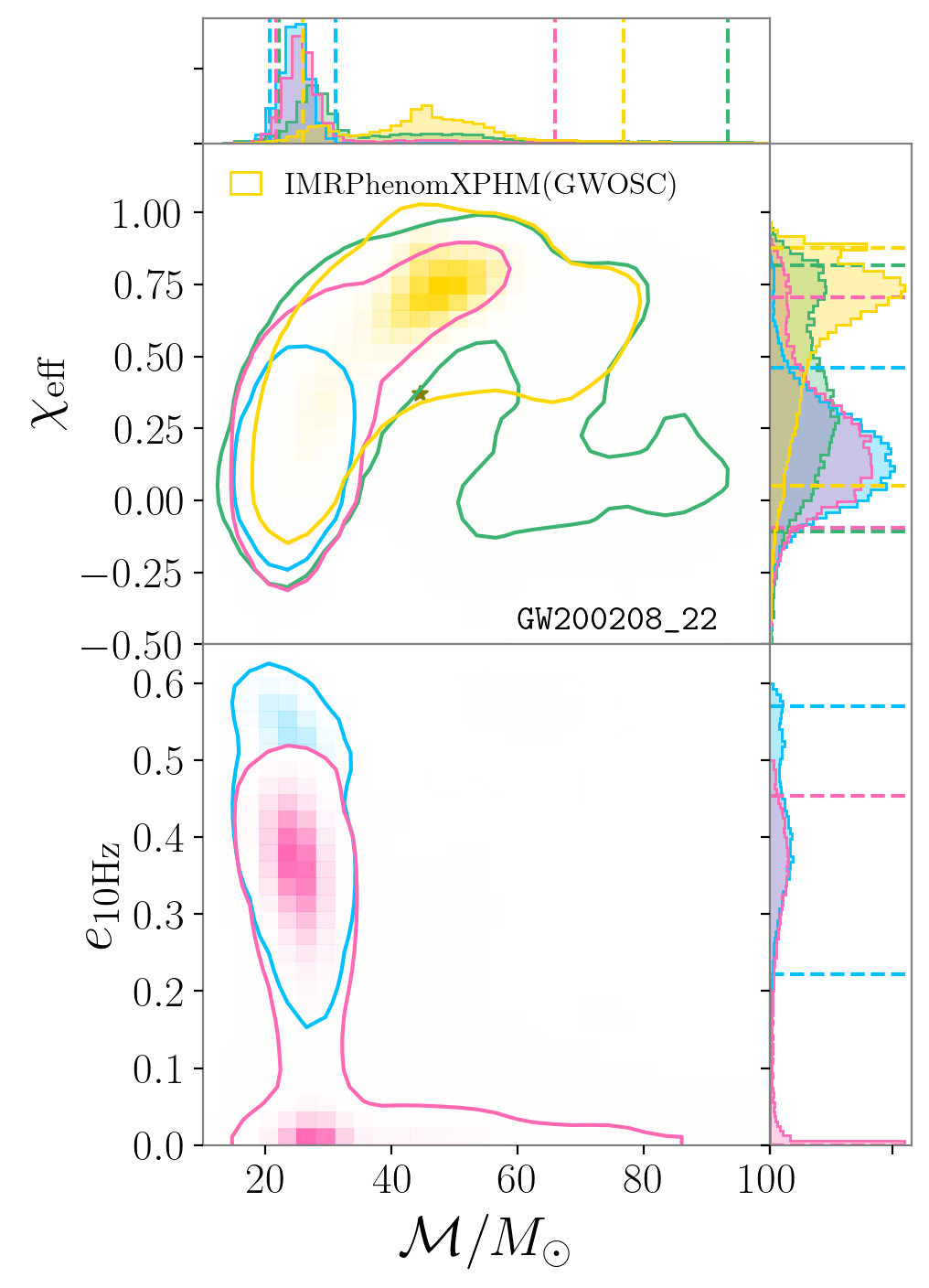}
    \caption{ Marginalized 2D and 1D posterior distributions of GW190701 (\textit{left}), GW190929 (\textit{middle}), and GW200208\_22 (\textit{right}) for the highest number of live points runs described in Tab. \ref{tab:GW_pes}. For each GW event we show the chirp mass and effective spin (\textit{top row}), and chirp mass and eccentricity(\textit{bottom row}) parameters;  for the QC model \phTHM (\textit{green}), and the eccentric \phTE model using both a uniform (Uni, \textit{blue}) and a log-uniform (LogUni, \textit{magenta}) priors in eccentricity. We include as a dark green star the median value of the posterior distribution obtained by \phTPHM. 
    All parameters are measured at a reference frequency of $f_{\mathrm{ref}}=10$ Hz.}
    \label{fig:GWevents}
\end{figure*}
GW200208\_22 is another event with evidence for the eccentric hypothesis ~\cite{Gupte:2024jfe, Romero-Shaw:2022xko}. For this analysis, we increase the upper bound of the eccentricity prior to $e_{\max}=0.65$ to prevent the posteriors from railing. For this high value of eccentricity, the underlying eccentricity expansions in the model degrade its accuracy~\cite{Planas:2025feq}. 
However, since the waveform starts at an orbit averaged $(2,2)$-frequency of 10Hz to ensure accurate higher-mode content, and the likelihood integration begins at 20Hz, the eccentricity is expected to have substantially decayed by 20Hz. As a result, any inaccuracies due to the eccentricity expansions are negligible at this frequency.
The runs for this event are summarized in Tab.~\ref{tab:GW_pes}. 
We perform PE runs with \phTHM, \phTPHM, and \phTE with both uniform and log-uniform priors, employing the default \texttt{bilby} settings specified in Sec.~\ref{subsec:PEbasis}. 
Furthermore, we repeat the \phTHM and \phTE runs with an increased number of live points (\texttt{nlive=2000}) to assess  stochastic sampler systematics in the multi-modalities observed in the QC parameters.

The high-resolution runs are shown in the third column of Fig.\ref{fig:GWevents}, together with those of \phXPHM from the GWTC-3 release~\cite{gwosc3, gwtc3}. This figure presents the 2D and 1D posterior distributions for the chirp mass, effective spin, and eccentricity --each evaluated at the reference frequency of 10 Hz-- for the three remaining special events, with one event per column. Notably, in these plots, we display the eccentricity as defined within the model rather than the GW eccentricity, which is listed in Tab.~\ref{tab:GW_pes}. This choice is motivated by the degradation of the waveform at higher eccentricities at 10 Hz, which prevents \texttt{gw\_eccentricity} from functioning correctly.
Since some cycles before the reference frequency are required to obtain the GW eccentricity values, we are evaluating the PN expansions at $e \sim 0.65-0.7$, which leads to non-physical behavior in the waveform.
However, since these values do not directly enter the waveform, we consider it more informative to show the full posterior, providing a clearer understanding of the range of eccentricities relevant at 20 Hz.

According to the values from the public samples in the GWTC-2.1 catalog~\cite{gwtc21}, using the \phXPHM model (included in Tab.~\ref{tab:GW_pes}), GW200208\_22 is identified as a BBH with total mass $M = 174^{+100}_{-106}$, effective spin $\chieff = 0.67^{+0.21}_{-0.62}$, and a relatively low SNR of $7.3^{+1.1}_{-2.0}$. 
As can be seen in Fig.~\ref{fig:GWevents}, the \phXPHM posteriors are not well constrained due to the low SNR. Thus, we study the use of more demanding sampler settings in \texttt{bilby}, both for the \phTHM and \phTE runs. 
We perform extra runs by increasing the number of live points to \texttt{nlive=2000} to reduce the multimodalities in the chirp mass and the eccentricity posteriors observed in the \texttt{nlive=1000} \phTE and \phTHM runs.

The uniform \phTE runs indicate a strong preference for the eccentric hypothesis over the QC one, with Bayes factors of $\log_{10}\mathcal{B}_{\mathrm{E/QC}}=0.96^{+0.12}_{-0.12}$ and $\log_{10}\mathcal{B}_{\mathrm{E/QC}}=1.14^{+0.08}_{-0.08}$ for the low- and high- resolution runs, respectively. These factors suggest support for elliptical orbits, with inferred GW eccentricities of $e^{\mathrm{GW}}_{10\rm{Hz}}=0.33^{+0.05}_{-0.16}$ and $e^{\mathrm{GW}}_{10\rm{Hz}}=0.34^{+0.08}_{-0.18}$.
However, this preference is reduced when using a log-uniform prior, yielding Bayes factors of $\log_{10}\mathcal{B}_{\mathrm{E/QC}}=0.09^{+0.11}_{-0.11}$ and $\log_{10}\mathcal{B}_{\mathrm{E/QC}}=0.49^{+0.08}_{-0.08}$, with corresponding GW eccentricities of $e^{\mathrm{GW}}_{10\rm{Hz}}=0.31^{+0.09}_{-0.30}$ and $e^{\mathrm{GW}}_{10\rm{Hz}}=0.29^{+0.09}_{-0.28}$, which diminish the evidence of orbital eccentricity in this event.
Nonetheless, as shown in Figs.~\ref{fig:posteriors} and ~\ref{fig:GWevents}, the log-uniform run recovers some support at the peak of the uniform-run distribution. Combined with the increase in the Bayes factor at higher resolution, this suggests that the lower preference observed in this case may be influenced by the event’s low SNR. In particular, the imposition of a prior that strongly favors non-eccentricity may lead to a preference for the QC hypothesis without necessarily providing strong evidence against the eccentric scenario.

We highlight that the inclusion of eccentricity can improve the measurement of binary parameters. As shown in the right column of Fig.\ref{fig:GWevents}, masses, spins, and extrinsic parameters like the luminosity distance (see Tab.\ref{tab:GW_pes} for details), are more tightly constrained when using the eccentric model compared to the QC model. 
Additionally, the masses inferred with the eccentric model are significantly lower than those obtained with the QC models, while the spins tend to concentrate closer to 0 --consistent with ensuring the same waveform duration.
While \phXPHM estimates a total mass of $\sim 174M_{\odot}$, \phTE infers $\sim 65M_{\odot}$, substantially reducing the total mass and allowing more waveform cycles to fall within the detector’s frequency band, making it a more favorable system for eccentricity detection.
These differences in binary parameter recovery underscore the importance of incorporating eccentric models in routine PE studies.

The evidence for eccentricity in this event is less significant due to its low SNR, which makes it more challenging to identify eccentric signatures, particularly when using a log-uniform prior in eccentricity. However, given the obtained Bayes factors, we cannot rule out the presence of eccentric features. 
In contrast, precession signatures appear disfavored, as indicated by the Bayes factor comparing the eccentric nonprecessing-spin hypothesis to the QC precessing-spin one:  $\log_{10}\mathcal{B}_{\mathrm{P/QC}}=-0.26^{+0.10}_{-0.10}$.

\begin{figure*}
    \centering
    \includegraphics[width=\linewidth]{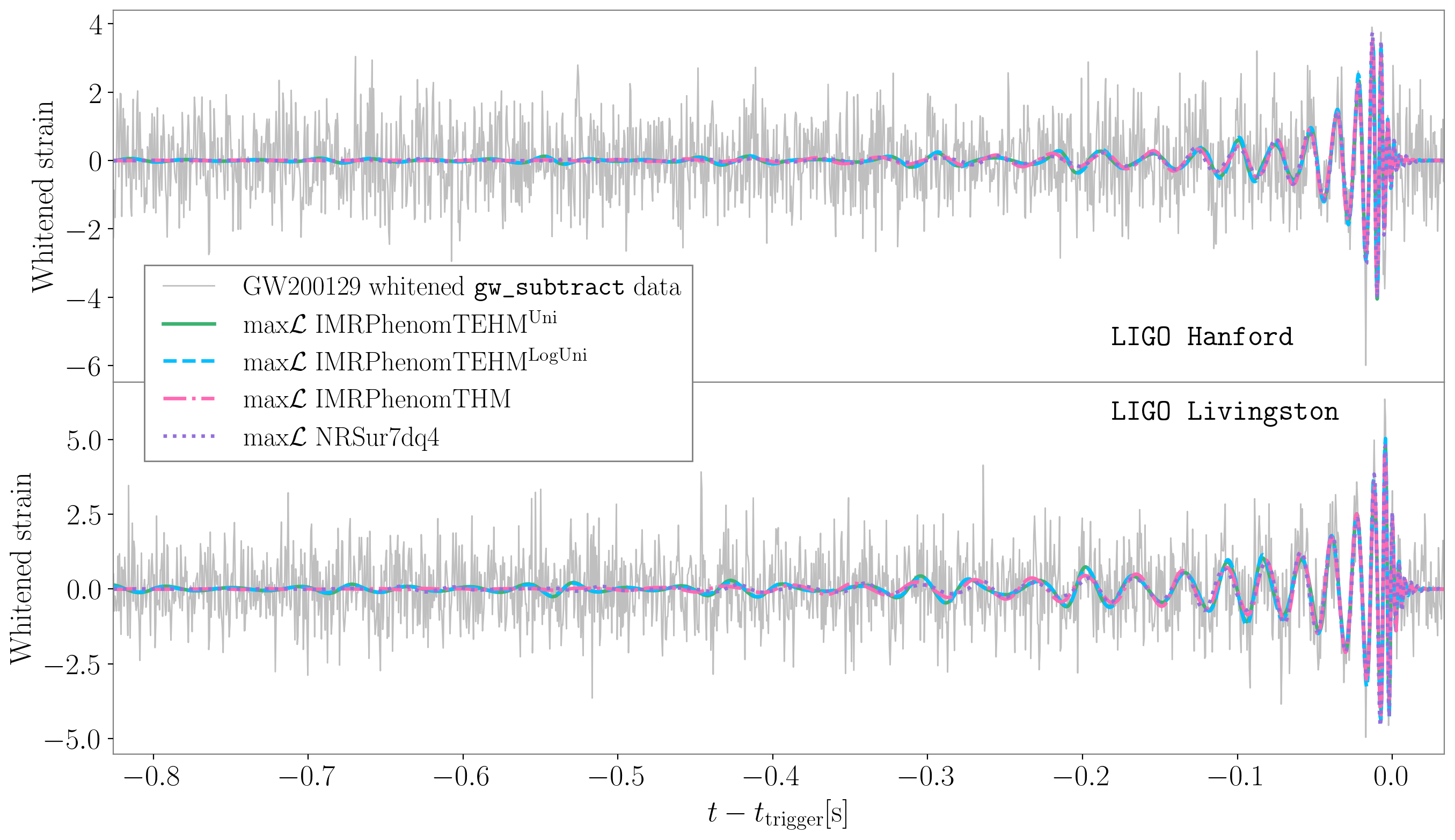}
    \includegraphics[width=\linewidth]{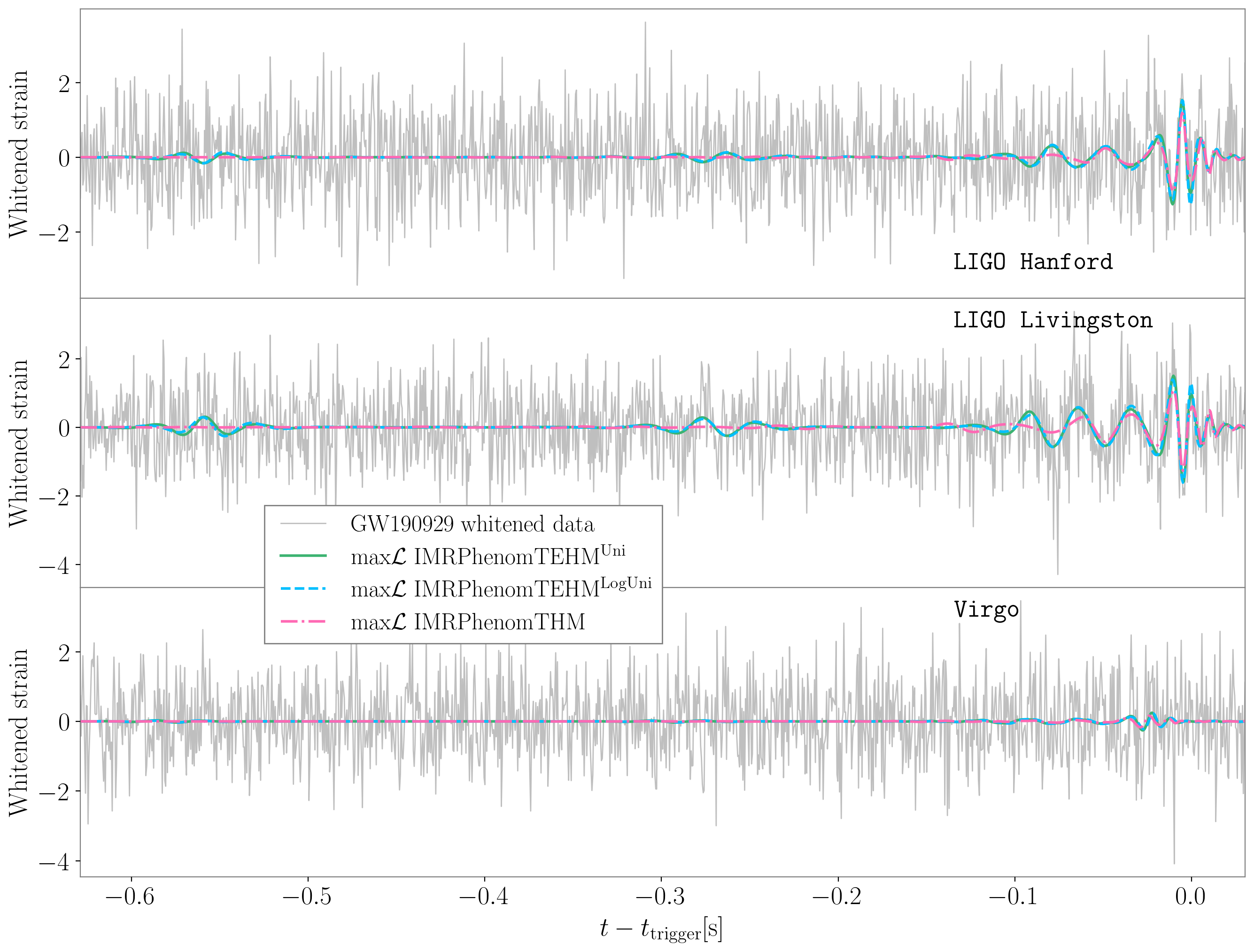}
    \caption{Whitened strain data (gray) and maximum likelihood waveform reconstructions for two events. \textit{Top}: GW200129 using whitened data from the \texttt{gw\_subtract} glitch mitigation technique. \textit{Bottom}: GW190929. Waveforms correspond to maximum likelihood configurations from different models: eccentric model \phTE with uniform (solid green) and log-uniform (dashed cyan) priors, the QC aligned-spin model \phTHM (dash-dotted magenta), and for GW200129, the \NRSur model (dotted purple).}
    \label{fig:waveforms}
\end{figure*}

\subsection{High mass events: GW190701 \& GW190929 }\label{subsec:highmass}

The two additional events that require further investigation are GW190701 and GW190929. These events have both been identified as non-spinning BBHs, with relatively high total masses: specifically, we find with \phTHM $M = 127.74^{+14.41}_{-13.52}M_{\odot}$ for GW190701 and $M = 141.64^{+23.26}_{-20.72}M_{\odot}$ for GW190929. Additionally, both events show a relatively low SNR of $11.38^{+0.13}_{-0.25}$ for GW190701 and $ 9.96^{+0.40}_{-0.52}$ for GW190929, when using the \phTHM model.
Both events show slight preference for the eccentric hypothesis with log-10 Bayes factors ranging from $0.09^{+0.10}_{-0.10}$ to $0.23^{+0.13}_{-0.13}$ 
for GW190701, and from $0.19^{+0.09}_{-0.09}$ to $0.56^{+0.11}_{-0.11}$ for GW190929, depending on the eccentricity prior considered.
The runs for these events are summarized in Tab.~\ref{tab:GW_pes}, and the higher resolution runs for \phTHM and \phTE with both eccentric priors are presented in the first and second columns of Fig.~\ref{fig:GWevents}.

GW190701 was first identified as a candidate for eccentricity in Ref.~\cite{Gupte:2024jfe} using \seobnrvfore \cite{Ramos-Buades:2021adz}, reporting a GW eccentricity of $e^{\rm GW}_{10Hz} = 0.35^{+0.32}_{-0.11}$ and log-10 Bayes factors of 3.0 and 2.61 for uniform and log-uniform eccentricity priors, respectively. 
The main difficulty in assessing evidence of eccentricity for this event is the very high total mass inferred from eccentric models (see Tab.~\ref{tab:GW_pes} for details), which implies that only a few cycles of the signal enter the detector's sensitive band, primarily capturing the merger-ringdown phase. As highlighted in our reanalysis of GW190521~\cite{Planas:2025feq}, making strong eccentricity claims in such cases is problematic, given the underlying assumption of circularization at merger in current eccentric models. Furthermore, GW190701 contains a known glitch, which we do not remove to be consistent with the data used in the GWTC-2.1 catalog~\cite{gwosc12,gwtc21}.

To our knowledge, no other references have reported signs of eccentricity in GW190929. However, in our study, this event exhibits the strongest support for eccentricity among the high-mass events, even when using a log-uniform prior. Despite this, the usual caveats associated with high total mass and low SNR must be carefully considered when interpreting these results.

As for GW200208\_22, we increase the upper bound of the eccentricity prior to $e_{\max}=0.65$ to prevent posterior railing. 
Our results using a uniform prior in eccentricity for both events suggest support across a wide range of eccentricities, with a preference for higher values. However, this support is largely suppressed when using a log-uniform prior (see Fig.~\ref{fig:posteriors} and Tab.~\ref{tab:GW_pes}). 
Nevertheless, the eccentricity posteriors from the log-uniform prior runs still show some support in the most probable region of the uniform prior eccentricity posteriors. To assess the impact of sampler systematics in the observed multimodalities we perform higher-resolution runs with an increased number of live points (\texttt{nlive=2000}).
The results of these high-resolution runs are shown in the first two panels of Fig.~\ref{fig:GWevents}. 
As seen in Tab.~\ref{tab:GW_pes} and the violin plots in Fig.~\ref{fig:posteriors}, the choice of sampler settings does not introduce significant differences in this case.
For both events, we find a similar pattern: while the uniform prior produces a broad posterior distribution in eccentricity, spanning from 0 to the upper bound of 0.65 with a peak around 0.5, the log-uniform prior significantly increases support for the QC hypothesis. However, it still retains some support for the mode around $e_{10\rm{Hz}}\sim0.5$.

An important consideration, already discussed in our reanalysis of GW190521~\cite{Planas:2025feq}, is that \phTE allows waveform generation for any reference frequency contained in the underlying eccentric dynamics.  This is not the case for other eccentric models, such as \seobnrvfore and \seobnrvfivee, which we do not have the ability to specify a reference frequency distinct from the starting frequency and impose constraints on waveform generation at small binary separations.

The observed multimodalities and the high total mass of these events complicates obtaining clear evidence of eccentricity, even if the Bayes factors slightly favor the eccentric hypothesis. Overall, high-mass events pose a challenge as only a few cycles before merger remain in the detector's sensitive band while our model --as well as all state-of-the-art eccentric models \cite{Nagar:2021xnh,Ramos-Buades:2021adz,Paul:2024ujx, Gamboa:2024hli,Gamba:2024cvy, Nagar_2024, nagar2025}-- assumes circularity at merger.
In the bottom panel of Fig.~\ref{fig:waveforms}, we show the whitened data for GW190929, overlaid with the maximum likelihood waveforms from the two eccentric \phTE runs (with uniform and log-uniform eccentricity priors), along with the aligned-spin quasi-circular run using \phTHM. As for GW200129, the eccentric reconstructions show remarkable agreement with each other. In this case, they differ from the QC waveform even at merger, although all reconstructions yield a similar ringdown. As seen from the comparison, it is difficult to definitively favor one waveform over another, particularly given the impact of tapering and model-specific transition treatments near merger for such short signals.
Therefore, in high-mass events, 
extra caution is needed to ensure that the evidence of eccentricity is not caused by spurious effects in the waveform due to the specifics on which different models transition from inspiral to merger. We leave for future work the incorporation of eccentric effects in the merger and ringdown of \phTE.

\section{Conclusions}\label{sec:conclusions}
In this work, we analyze 17 GW events from BBHs that have been identified in the literature as particularly interesting due to their support for dynamical formation channels. For this analysis, we use the new phenomenological time-domain multipolar waveform model \phTE~\cite{Planas:2025feq}, which incorporates two eccentric parameters: eccentricity and mean anomaly. This model builds upon the QC aligned-spin \phTHM model and includes up to 3PN eccentric corrections to both the dynamics and waveform multipoles.

Previous eccentric studies have relied on resampling techniques, iterative fitting, or machine learning algorithms~\cite{Romero-Shaw:2019itr,Iglesias:2022xfc,Gupte:2024jfe}\footnote{Note that Ref. \cite{Gupte:2024jfe} employs machine learning algorithms to analyze the bulk of GW events, as well as the \texttt{parallel Bilby} \cite{parallel_bilby_paper} code for a subset of events.}. In contrast, this work allows, for the first time, the routine use of an eccentric waveform model with the standard analysis tools of the LVK Collaboration due to its high computational efficiency.
We conduct analysis with the aligned-spin QC model \phTHM and the eccentric \phTE, employing both uniform and log-uniform priors in eccentricity for all events. For the four events requiring further investigations, we perform additional runs, varying data and sampler settings, as well as runs with the precessing models \phTPHM and \NRSur. To complement our results, we provide a data release~\cite{zenodo_release} containing all the posterior samples generated in this work.

Our results provide valuable insights into the measurement of eccentricity in PE studies and their interpretation, highlighting the need to include eccentricity in PE studies to prevent biased in inferred parameters.
In particular, we find support for eccentricity in four events: GW190701, GW190929, GW200129, and GW200208\_22. Given these indications and thanks to the computational efficiency of the model, we performed additional studies to assess the robustness of our findings and to explore the impact of glitch subtraction and waveform systematics.

Our analysis of GW200129 provides evidence for orbital eccentricity, as this feature is consistently supported across all eccentricity priors and glitch mitigation techniques. 
This reinforces the interpretation of a dynamical formation channel, as suggested by previous studies~\cite{Hannam_2022,Gupte:2024jfe}. The non-deglitched \texttt{GWOSC} data yield the highest Bayes factors for eccentricity. Using the \texttt{gw\_subtract} mitigation method, widely adopted in LVK analyses, we continue to find robust support for eccentricity, albeit slightly lower than in the uncorrected data. 
The \texttt{BayesWave} mitigation, particularly the \texttt{BayesA} draw, exhibits the least support for dynamical features (both precession and eccentricity), highlighting the impact of the glitch treatment on parameter estimation. However, even with this mitigation, the eccentric hypothesis remains favored over both the aligned-spin and precessing QC hypotheses. 
Given the high SNR and total mass of GW200129, which make it a good candidate for detecting eccentricity in ground-based detectors, our findings support the presence of orbital eccentricity. While we cannot entirely rule out precession, our results suggest that eccentricity is the dominant feature. 
One of the main limitations in our analysis stems from the lack of precessing-spin effects in the \phTE model, which is crucial to distinguish features from spin precession and eccentricity. Some first IMR eccentric precessing-spin models have recently emerged,  see e.g.~Refs.~\cite{Liu_2023, Gamba:2024cvy, Albanesi_2025}, and we plan to extend \phTE to the fully generic parameter space including both eccentricity and spin precession in future work.

Our analysis of GW200208\_22 also provides some support for the eccentric hypothesis, although the conclusions are not as definitive as those for GW200129. The evidence for eccentricity is strengthened by using a uniform prior, yielding a log Bayes factor of $1.14^{+0.08}_{-0.08}$. However, the log-uniform prior provides minor support in favor of eccentricity with a log-10 Bayes factors of $0.49^{+0.08}_{-0.08}$. 
In both cases the results show that the eccentric hypothesis remains favored over QC precessing-spin and QC aligned-spin models, particularly in the higher-resolution PE runs.
However, the relatively low SNR of the event complicate the analysis, and make robust conclusions more challenging.

Finally, GW190701 and GW190929 are two high-mass BBH events that show minor preference for the eccentricity hypothesis, though their characterization remains challenging due to their large total masses and low signal-to-noise ratios. The high inferred total masses imply that only the final cycles before merger are observed, where eccentric waveform models, including \phTE, typically assume circularization. 
For GW190929, we find stronger support for eccentricity among the high-mass events, although systematic uncertainties remain. We find that the uniform prior favors the eccentricity scenario (log-10 Bayes factors of $0.56^{+0.09}_{-0.09}$), while the log-uniform prior suppresses this support (log-10 Bayes factors of $0.19^{+0.09}_{-0.09}$). 
GW190701 was previously identified as an eccentric candidate in Ref.~\cite{Gupte:2024jfe}, with  $\log_{10}\mathcal{B}_{\mathrm{E/QC}}$ values of 3.0 and 2.11 for uniform and log-uniform priors, respectively.
While our log-10 Bayes factors also show a slight preference for the eccentric hypothesis over the aligned-spin QC scenario, they oscillate depending on the sampler settings and choice or eccentricity priors, ranging between $\sim 0.05-0.24$. These low values, combined with the presence of a known glitch, prevent strong conclusions and do not support the claim obtained with \seobnrvfore.
Even though our results suggest possible eccentricity signatures in these events, the short duration of the signals and the limitations of the eccentric modeling near merger prevent strong claims, highlighting the need for improved waveform models that include eccentric corrections up to merger.

Our findings emphasize the crucial need for eccentric waveform models in GW PE analyses. 
This study identifies significant signs of eccentricity in 2 out of approximately 90 events detected by the LVK Collaboration, suggesting that future observing runs may reveal a comparable or even higher occurrence rate.
Omitting eccentricity leads to systematic biases in key parameters such as chirp mass and effective spin, which arise from the waveform systematics introduced when eccentricity is ignored. Since eccentric waveforms are typically shorter than their QC counterparts, the total mass is underestimated to match the observed signal duration. Similarly, we observe a strong interplay between eccentricity and the effective spin parameter, as previously noted in Fig.~8 of Ref.~\cite{Planas:2025feq}. This interplay underscores the necessity of developing a fully generic, spin-precessing eccentric model suitable for comprehensive GW analyses.
Moreover, eccentricity provides valuable insights into the dynamical formation channels of binary systems, and as such, improving waveform models for generic orbits is essential. These models must not only be accurate, but also computationally efficient to enable their systematic application across large datasets.

Finally, the high computational efficiency introduced by \phTE enables us to perform PE for low total-mass events with very long duration signals. 
This opens the door to a more detailed investigation of recent claims of eccentricity in GW200105~\cite{Morras:2025xfu}, one of the NSBH events reported in Ref.~\cite{LIGOScientific:2021qlt}.
In an upcoming paper, we analyze this event using, for the first time, a fully eccentric IMR waveform model with standard stochastic sampling techniques~\cite{Lluc_NSBH_2025}.

\section*{Acknowledgements}

The authors would like to thank Nihar Gupte for the LSC Publication \& Presentation Committee review of this manuscript.
We thankfully acknowledge the computer resources (MN5 Supercomputer), technical expertise and assistance provided by Barcelona Supercomputing Center (BSC)  through funding from the Red Española de Supercomputación (RES) (AECT-2024-3-0027); and the computer resources (Picasso Supercomputer), technical expertise and assistance provided by the SCBI (Supercomputing and Bioinformatics) center of the University of Málaga (AECT-2025-1-0035).
This research has made use of data or software obtained from the Gravitational Wave Open Science Center (gwosc.org), a service of the LIGO Scientific Collaboration, the Virgo Collaboration, and KAGRA.
This material is based upon work supported by NSF's LIGO Laboratory which is a major facility fully funded by the National Science Foundation.
LIGO is funded by the U.S. National Science Foundation. Virgo is funded by the French Centre National de Recherche Scientifique (CNRS), the Italian Istituto Nazionale della Fisica Nucleare (INFN) and the Dutch Nikhef, with contributions by Polish and Hungarian institutes.

Maria de Lluc Planas is supported by the Spanish Ministry of Universities via an FPU doctoral grant (FPU20/05577).
A. Ramos-Buades is supported by the Veni research programme which is (partly) financed by the Dutch Research Council (NWO) under the grant VI.Veni.222.396; acknowledges support from the Spanish Agencia Estatal de Investigación grant PID2022-138626NB-I00 funded by MICIU/AEI/10.13039/501100011033 and the ERDF/EU; is supported by the Spanish Ministerio de Ciencia, Innovación y Universidades (Beatriz Galindo, BG23/00056) and co-financed by UIB.
CG is supported by the Swiss National Science Foundation (SNSF) Ambizione grant PZ00P2\_223711.
This work was supported by the Universitat de les Illes Balears (UIB); the Spanish Agencia Estatal de Investigación grants PID2022-138626NB-I00, PID2019-106416GB-I00, RED2022-134204-E, RED2022-134411-T, funded by MCIN/AEI/10.13039/501100011033; the MCIN with funding from the European Union NextGenerationEU/PRTR (PRTR-C17.I1); Comunitat Autonòma de les Illes Balears through the Direcció General de Recerca, Innovació I Transformació Digital with funds from the Tourist Stay Tax Law (PDR2020/11 - ITS2017-006), the Conselleria d’Economia, Hisenda i Innovació grant numbers SINCO2022/18146 and SINCO2022/6719, co-financed by the European Union and FEDER Operational Program 2021-2027 of the Balearic Islands; the “ERDF A way of making Europe”.






\let\c\Originalcdefinition %
\let\d\Originalddefinition %
\let\i\Originalidefinition

\bibliography{bib_TEHM}

\end{document}

%% file: UsePackages.tex
\usepackage[T3,T1]{fontenc}
\usepackage{mathrsfs} 
\usepackage{bm} 
\makeatletter
\@ifclassloaded{beamer}
  { 
    \typeout{UsePackages: Detected beamer}
    \usepackage{tgheros}
    
  }
  { 
    \typeout{UsePackages: Did not detect beamer}
    \ifx\asybeamer\undefined 
    \typeout{UsePackages: Detected article}
    \usepackage[varg]{txfonts} 
    \usepackage{tgtermes} 
    \else 
    \typeout{Fonts: Detected asy for beamer}
    \usepackage{tgheros}
    
    \fi
  }
\usepackage{microtype} 
\def\MT@register@subst@font{
  \MT@exp@one@n\MT@in@clist\font@name\MT@font@list
  \ifMT@inlist@\else\xdef\MT@font@list{\MT@font@list\font@name,}\fi}
\makeatother

\usepackage{graphicx} %
\usepackage[x11names,svgnames,rgb]{xcolor} %
\usepackage{xspace}
\usepackage{braket}
\usepackage{accents}
\usepackage{siunitx}
\usepackage{mathtools}
\DeclareSymbolFontAlphabet{\mathrm}{operators}

\definecolor{CiteColor}{rgb}{0.18039, 0.18824, 0.57255}
\definecolor{UrlColor} {rgb}{0.741, 0.173, 0.000}
\definecolor{DarkUrlColor} {rgb}{0.500, 0.110, 0.000}
\definecolor{LinkColor}{rgb}{0.25098, 0.47843, 0.04706}

\makeatletter %
\newcommand{\ShowFont}{%
  \typeout{The main font is \f@encoding \space \f@family \space %
    \f@series \space \f@shape \space at \f@size pt.}%
  \typeout{The math font sizes are \tf@size pt (main), \sf@size pt %
    (script), and \ssf@size pt (scriptscript).}%
  \typeout{The linewidth is \the\linewidth}} %
\makeatother %

\usepackage{xargs}
